# Integrating electronic structure into generative modeling of inorganic materials

*Junkil Park[1,2], Junyoung Choi[1] and Yousung Jung[1,2,3*]*

[1] Department of Chemical and Biological Engineering, Seoul National University, 1 Gwanak-ro, Gwanak-gu, Seoul 08826, Korea

[2] Institute of Engineering Research, Seoul National University, 1 Gwanak-ro, Gwanka-gu, Seoul 08826, Korea

[3] Institute of Chemical Processes, Seoul National University, 1 Gwanak-ro, Gwanka-gu, Seoul 08826, Korea

# Abstract

Recent advances in generative models have introduced a new paradigm for the inverse design of inorganic materials, enabling the discovery of new crystalline structures with desired properties. However, existing generative models focus solely on structural aspects of materials during generation, while overlooking the underlying electronic behavior that fundamentally governs materials' stability and functionality. In this work, we present ChargeDIFF, the first generative model for inorganic materials that explicitly incorporates electronic structure into the generation process. Specifically, ChargeDIFF leverages charge density, a direct spatial representation of a material's electronic structure, as an additional modality for generation. ChargeDIFF demonstrates exceptional performance in both unconditional and conditional generation tasks compared to baseline models, with ablation studies revealing that this improvement is directly due to its ability to capture the material's electronic structure during generation. Moreover, the ability to control charge density during generation allows ChargeDIFF to introduce a novel inverse design method based on three-dimensional charge density, illustrating the potential to generate lithium-ion battery cathode materials with desired ion migration pathways, as validated by physics-based simulations. By highlighting the importance of accounting for electronic characteristics during material generation, ChargeDIFF offers new possibilities in the generative design of stable and functional materials.

# Introduction

The development of materials with desired properties has long been the ultimate goal in the field of materials science, which drives innovation across a wide range of industries[1-3]. In particular, inorganic materials have been at the center of attention for their structural and chemical versatility, which underpins key applications in energy storage[4,5], catalysis[6,7] and semiconductors[8,9]. Traditionally, discovery of such functional materials has relied heavily on experimental trial-and-error processes guided by the intuition of researchers, which is constrained by the limited number of target structures[10]. Recent advances in computational chemistry and machine learning have transformed this paradigm, enabling the virtual high-throughput screening of hundreds of thousands of candidate structures within the database[11-13]. Although this high-throughput approach has accelerated progress across various applications, it remains inherently confined by the limited search space defined by known materials, representing only a small fraction of the vast and largely unexplored chemical space of crystalline structures[14].

Over the past decade, rapid advances in generative models have established a new paradigm in materials discovery[15,16]. Generative models learn the intrinsic distribution of materials data, and enable the generation of new materials without being restricted to the search space defined by existing databases. Various generative model architectures, such as variational autoencoder (VAE)[17,18] and generative adversarial network (GAN)[19,20], have been applied to inorganic materials generation, successfully demonstrating a proof of concept in the early stages of adopting generative models in this field. Most recently, diffusion models have emerged as the mainstream approach due to their superior generation performance, and a number of diffusion-based models for inorganic materials generation have been proposed to date[21-25]. In particular, MatterGen[23] has garnered significant attention by achieving experimental validation of inversely designed materials, bringing generative models one step closer to practical applicability.

Reflecting on the progress made so far, most reported diffusion models for inorganic materials generation, with the exception of few unconventional approaches[26], consider only the structural components of materials during the generation process[21-23,25,27]. Typically, materials are represented as a combination of atom types (**A**), atomic coordinates (**X**), and lattice parameters (**L**), which serve as both inputs and outputs within these models. We refer to this as the **AXL** representation in this paper. Although this representation provides a minimal basis for describing inorganic materials, it remains unclear whether it sufficiently captures the underlying electronic structure, which fundamentally governs materials' stability and functionality, during the generation process. Given that recent studies have primarily focused on expanding databases[23,25], refining algorithms[22,23], or incorporating new technical features[27] while adhering to the representational conventions established by earlier successful models (e.g., CDVAE[21]), revisiting the foundational question of how materials are represented during the generation process is timely and crucial for the future advancement of this field.

In this work, we propose ChargeDIFF, the first generative model for inorganic materials that explicitly considers electronic structure during the generation process. Specifically, ChargeDIFF is a diffusion model that considers valence electron density (conventionally and hereafter referred to as charge density, denoted as **C**), which is the direct spatial representation of a materials' electronic structure, as an additional modality for generation alongside the conventional **AXL** representations (Fig. 1a). We refer to this new method as the **AXLC** representation. We note that incorporating high-dimensional electronic structure data such as charge density as a generation modality has not been reported for inorganic materials generation. ChargeDIFF thus represents one of the first demonstrations of such modality expansion in materials generative modeling.

Building upon additional architectural features, including a vector quantized-variation autoencoder (VQ-VAE)[28], ChargeDIFF exhibited strong performance in unconditional generation task and various inverse design tasks compared to baseline models. Ablation studies showed that the performance improvements arise from the model's deeper understanding of the electronic characteristics of generated structures, enabled by the use of charge density. Moreover, the introduction of three-dimensional (3D) charge density data allows a novel generation approach based on inpainting technique[29], which we showcase by generating lithium-ion battery (LIB) cathode materials with user-desired ion migration pathways.

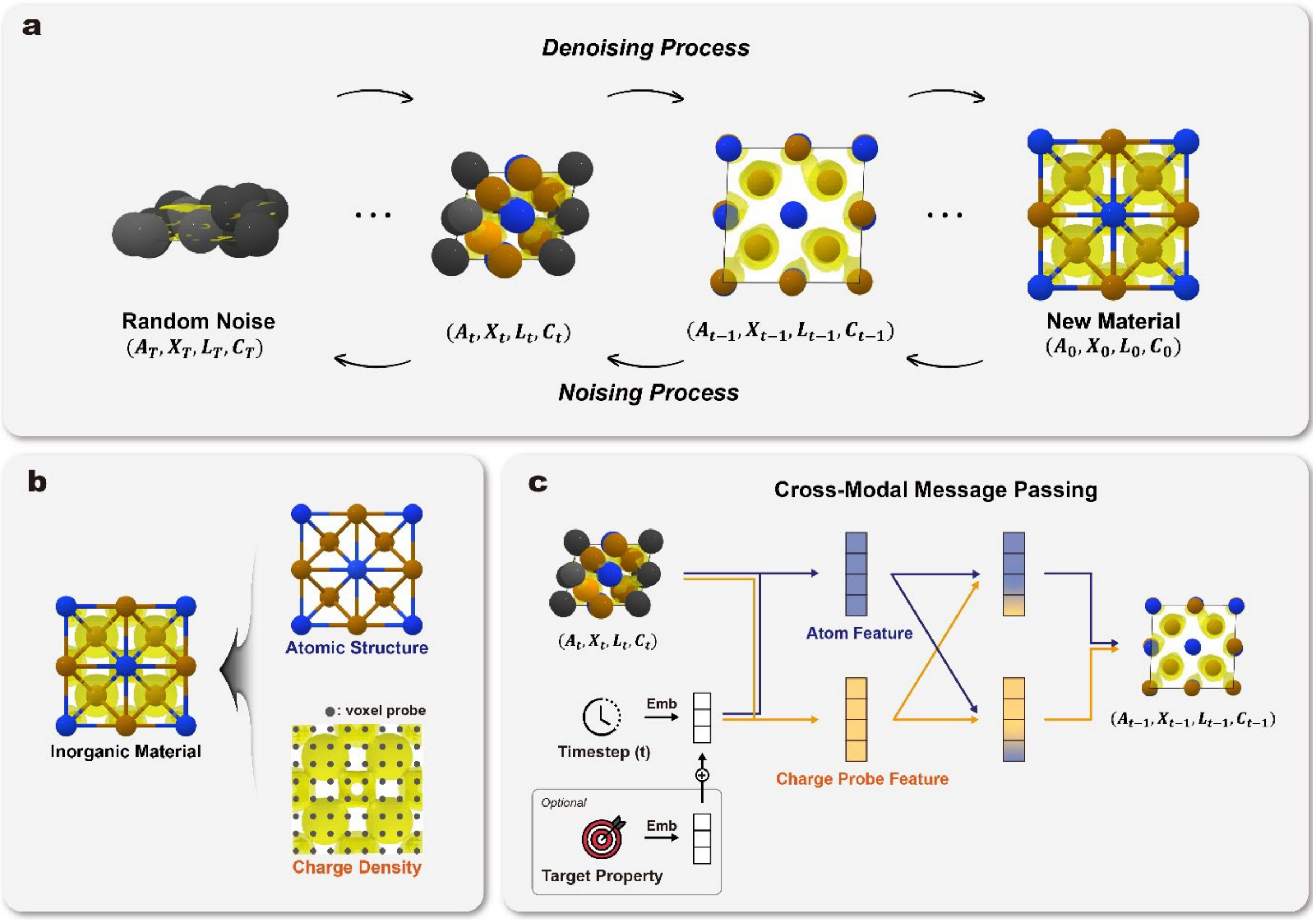

**Fig. 1. ChargeDIFF Overview. a.** A graphical illustration of the noising and denoising process in ChargeDIFF. Inorganic materials are represented as a combination of atom types (A), atomic coordinates (X), lattice parameters (L), and charge density (C), and these components undergo a joint diffusion process. **b.** Incorporation of charge density. Charge density data is incorporated into the representation of inorganic materials as voxel-based data, comprising a 3D arrangement of probes, each of which contains the charge density value at its corresponding spatial location. **c.** Cross-modal message passing during the denoising process. During the denoising process, atom node features and charge probe node features mutually influence each other to predict the material state at the previous timestep.

# Results

## Diffusion Process of Materials in ChargeDIFF

In the diffusion process of ChargeDIFF, an inorganic material is represented as a combination of atom types, atomic coordinates, lattice parameters, and charge density (i.e., **A**, **X**, **L**, and **C**), where the incorporation of charge density **C** is the new distinctive feature of ChargeDIFF compared with previous works (Fig 1a). Charge density is represented as a three-dimensional voxel, where charge density values at each location assigned to uniformly distributed probe points within the unit cell (Fig. 1b). This follows the standard representation of charge density in inorganic materials and is consistent with the file format used in the widely adopted VASP package[30] for density functional theory (DFT) calculations.

Each of the four components defining the material (**A**, **X**, **L**, and **C**) undergoes a joint diffusion process, where they are progressively corrupted into random noise during the noising process and this corruption is reversed during the denoising process. In ChargeDIFF, the diffusion process for each component is carefully formulated to reflect its inherent characteristics. Specifically, the diffusion process of discrete atom types is modeled using the Discrete Denoising Diffusion Probabilistic Model (D3PM)[31], while the continuous-valued lattice parameters and charge density are handled with the standard Denoising Diffusion Probabilistic Model (DDPM)[32]. For atomic coordinates, which are both cyclic and bounded due to periodic boundary conditions (PBCs), the diffusion process is formulated using score matching with a Wrapped Normal distribution[33]. A more detailed description of the diffusion processes in ChargeDIFF is provided in the Methods section.

Once the diffusion process is formulated, a denoising network is trained to predict the noise (or score) at each state, and is then applied iteratively during the denoising process to progressively refine randomly sampled noise into new structures. A central aspect of the ChargeDIFF workflow is how the newly incorporated charge density is processed in conjunction with other structural components within the denoising network. In ChargeDIFF, the denoising network is uniquely implemented as a heterogeneous graph neural network (GNN), in which atoms and charge density probes are represented as distinct types of nodes within the same graph. In particular, charge probe nodes behave differently during the denoising process, with their coordinates fixed at the grid points and only their features updated, unlike atom nodes, whose coordinates are also adjusted. The architecture is inspired by previous works that predicted charge density of a given structure at specific spatial query locations[34,35]. Within the denoising network, message passing between atom probe nodes and charge probe nodes (referred to as cross-modal message passing, Fig. 1c) is facilitated, enabling structural and electronic components to mutually influence each other during generation, eventually leads to the generation of electronically stable structures.

## Latent Diffusion Approach using VQ-VAE

For the MP-20-Charge dataset used in training (see Methods section for detail), the charge density data for each structure is prepared in a voxel resolution of 32×32×32. However, this requires the aforementioned denoising network to process $32^3$ probe nodes within the graph, which is computationally impractical. It is also possible to process the charge density to a lower resolution (e.g., 8×8×8) via interpolation, but it results in a substantial loss of fine-grained information and significantly degrades generation quality (further discussed in later sections). Consequently, an approach is required that reduces the computational demands while maintaining fine-grained electronic information as possible, ensuring the effective generation that accurately reflects the charge density.

A field that has faced similar computational challenges is high-resolution image generation[36,37]. In this domain, latent diffusion models[36], which operate by encoding high-resolution images into a lower-resolution latent space during generation, have proven effective and facilitated the creation of high-quality images. Inspired by this, we adopt a similar strategy for handling charge density data by mapping it to a lower-resolution latent space. To achieve this, a VQ-VAE is implemented within the framework, which is an essential architectural component that compresses continuous input data into quantized codebook tokens (Fig. 2a)[28]. In ChargeDIFF, the VQ-VAE encoder compresses the charge density resolution from 32×32×32 to 8×8×8, while preserving the rich information contained in the high-resolution voxels. This compression enables the denoising network to operate on a significantly reduced number of probes nodes, substantially improving the computational efficiency of the generation process (Supplementary Fig. S3).

Within the overall workflow, the VQ-VAE is first trained independently and then serving as a fixed encoder during the training of the denoising network. A detailed description of the VQ-VAE training process can be found in the Methods section. To verify that the trained VQ-VAE captures meaningful features and effectively preserves charge density information during the mapping process, the latent space of VQ-VAE was visualized using t-distributed stochastic neighbor embedding (t-SNE)[38] in Fig. 2b. The embedded latent charge density vectors corresponding to structures with the top 5 most frequently occurring space groups in the database were assessed, revealing that the charge densities belonging to the same space group cluster together within the latent space. A similar clustering pattern was also observed in the comparative visualization based on the metallicity of the structures (Supplementary Fig. S5). The results indicate that the trained VQ-VAE effectively captures both the geometric and chemical features of the charge density during the mapping process, allowing it to be handled much more efficiently through the latent diffusion approach in ChargeDIFF's material generation process.

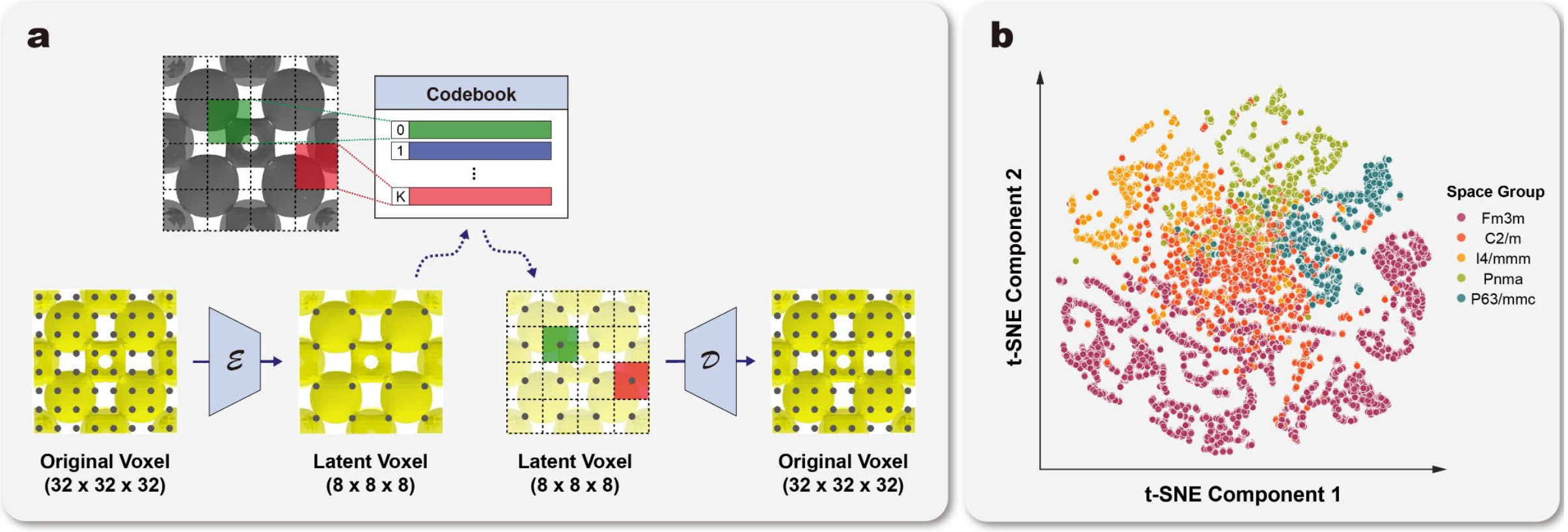

**Fig. 2. VQ-VAE for charge density compression and restoration. a.** The architecture of VQ-VAE implemented within ChargeDIFF. The VQ-VAE is composed of encoder and decoder, with a codebook used for vector quantization for mapping into a discrete latent space. Through the VQ-VAE, the charge density voxels are compressed into a lower-resolution, which significantly reduces the number of voxel probes that need to be processed during the diffusion process. **b.** A t-SNE plot of latent charge density data, depicting the top 5 most frequently occurring space groups within the training database.

**Electronic Structure-aware Generation of Inorganic Materials**

ChargeDIFF is capable of generating inorganic structures comprising a diverse set of elements across the periodic table (Supplementary Fig. S7), with several random samples shown in Fig. 3a. It can be observed that charge densities are generated alongside other structural components, and their visualizations qualitatively confirm they exhibit physically plausible distributions, forming approximately spherical shapes around each atom. This indicates that, during the material generation process, the charge density interacted with other structural components via cross-modal message passing, and the generated structures reflect the outcome of these mutual interactions.

To quantitatively evaluate the consistency between the generated charge density and the corresponding atomic structures, we recalculated the charge density from the generated structures using DFT and compared it with that generated by ChargeDIFF. The difference between the DFT-calculated and ChargeDIFF-generated charge densities is visualized in Fig. 3b. It can be seen that the two charge densities are largely consistent, and a quantitative evaluation over 1,000 generated structures indicates a mean square deviation of 2.2 % (see Methods section). This suggests that ChargeDIFF effectively captures the electronic structure during the generation process, resulting in self-consistent structures in which the charge density and atomic configuration are mutually compatible.

To assess the capability of ChargeDIFF in generating stable structures, the thermodynamical stability of 1,000 generated structures was assessed based on energy above hull values calculated using DFT. When evaluated against commonly used stability criterion of 0.1 eV/atom[23] (which is also occasionally described as metastable[25]), 54.4 % of the generated structures were found to be stable, confirming that ChargeDIFF can generate energetically stable structures (Fig. 3c). Currently, the most widely used metric for evaluating and comparing the performance of inorganic material generation models is the percentage of SUN (i.e., stable, unique, and new) structures among the generated structures. This metric not only reflects the generation of stable structures but also serves as a composite measure to evaluate whether the model can produce a wider variety of structures that were not seen during training (see Supplementary Information Section S3-2 for details). Evaluation of generative performance using this metric shows that ChargeDIFF achieves a high percentage of SUN structures, surpassing previously reported baselines (Fig. 3d) and thereby demonstrating its superior generation performance.

To investigate the origin of the outstanding generative performance of ChargeDIFF, an ablation study was conducted. A variant model in which the charge density was removed from the representation (thus matching conventional **AXL** representation), while all other components remained unchanged, denoted as ChargeDIFF (AXL), was evaluated for its material generation performance. As shown in Fig. 3d, ChargeDIFF (AXL) exhibited a SUN structure ratio of 21.3 %, which is considerably lower than that of original ChargeDIFF model (24.8 %) and, in some cases, lower than certain baseline models. These results provide direct evidence that the superior generation performance of ChargeDIFF arises from the incorporation of charge density, which captures the electronic structure information and thereby enables the generation of chemically stable structures.

In addition, to evaluate the effectiveness of the latent diffusion approach implemented via VQ-VAE within ChargeDIFF, an ablation study on the latent diffusion process was conducted. A variant model was trained in which the VQ-VAE was removed, directly using the lower resolution (8×8×8) charge density data without latent mapping (denoted as ChargeDIFF (w/o VQ-VAE)), and its generation performance was assessed. As shown in Supplementary Fig. S8, for structures generated by ChargeDIFF (w/o VQ-VAE), the deviation from DFT-calculated charge density increased to 20.6 %, and the SUN structure ratio decreased to 19.6 %. Notably, this SUN ratio is even lower than that of ChargeDIFF (AXL), which does not incorporate charge density at all. These findings demonstrate that incorporating latent diffusion through VQ-VAE is critical, and that the architectural refinements of ChargeDIFF being central to the improved generation performance.

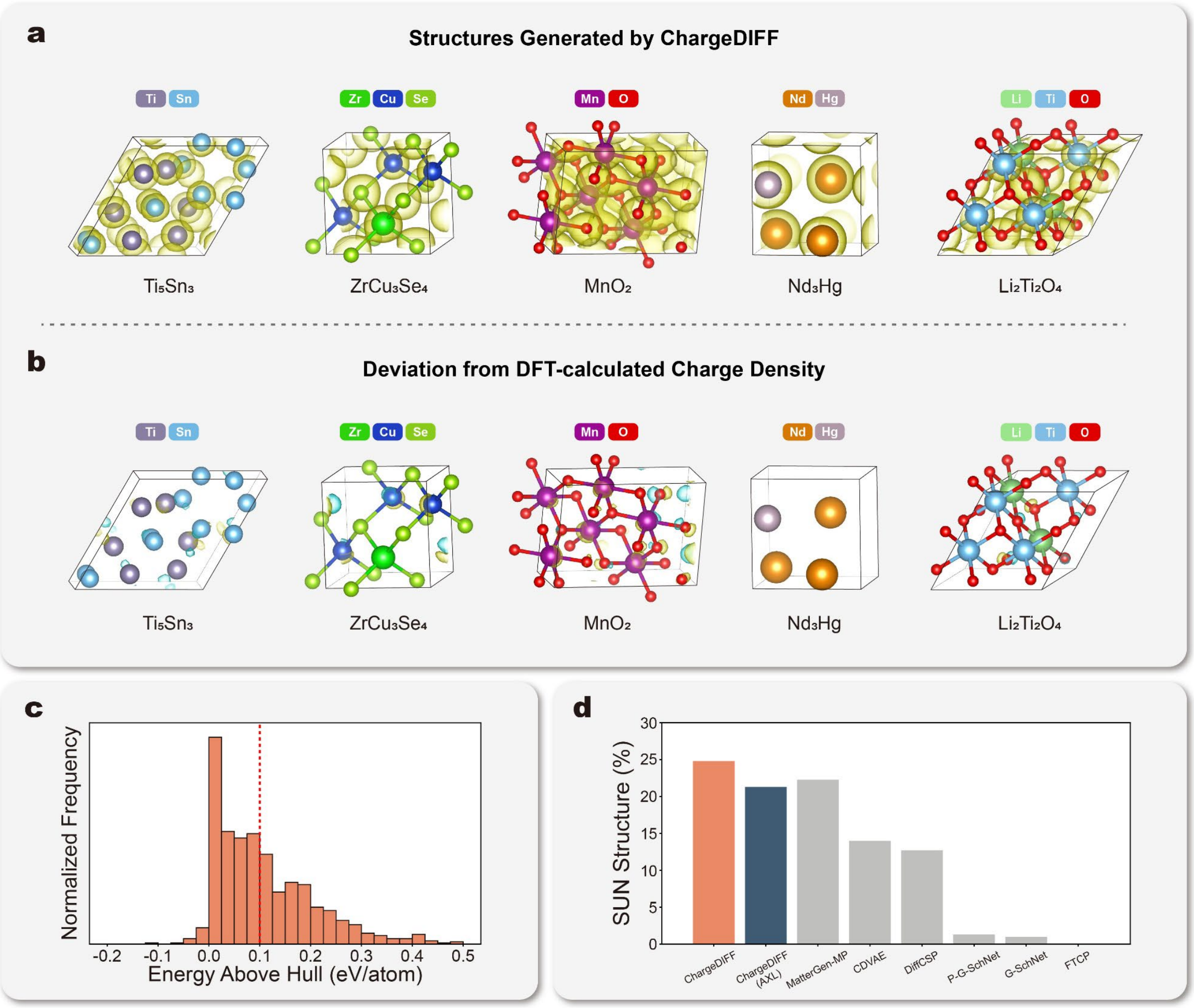

**Fig. 3. Unconditional Generation Results. a.** Generated sample structures from ChargeDIFF. Isosurfaces of the generated charge density at +0.5 e/Å$^3$ are visualized in yellow. **b.** Visualization of the difference between the generated charge density and the charge density recalculated via DFT from the generated structure. Isosurfaces of generated charge density at +0.5 e/Å$^3$ (yellow) and -0.5 e/Å$^3$ (blue) are visualized. **c.** Distribution of energy above hull values of 1,000 generated structures using ChargeDIFF. The stability of materials was evaluated using 0.1 eV/atom as the threshold. **d.** Comparison of generation performance in terms of the percentage of SUN structures. ChargeDIFF (orange) is compared with its ablated variant without charge density, ChargeDIFF (AXL) (blue), and several baseline models (gray), including MatterGen-MP[23], CDVAE[21], DiffCSP[22], P-G-SchNet, G-SchNet[39], and FTCP[18].

## Inverse Design of Materials with Desired Chemical Properties

Conditional generation of materials with desired chemical properties is an ultimate goal of generative modeling, and the capability of ChargeDIFF in generating materials with target chemical properties is assessed. ChargeDIFF utilizes a standard conditioning approach, in which the target property information is embedded and concatenated with the timestep embedding. The chemical property labels for model training were obtained from the values reported in the Materials Project database[40]. A detailed technical description of ChargeDIFF's conditional generation process is provided in Supplementary Information Section S4.

The target properties considered include bandgap, magnetic density, and crystal density, which encompass the material's electrical, magnetic, and structural characteristics, respectively. For each property, two different target values were considered, and the distributions of the corresponding property values for 500 generated structures are shown in Fig. 4a-c. It can be seen that each generation exhibited distinct distributions well aligned with the respective target values, implying that ChargeDIFF can selectively generate structures according to the specified property targets. An undesired peak near 0 eV appeared in the bandgap distribution, which is interpreted as arising from the severe imbalance in the training data, as the majority of structures exhibit near-zero bandgaps (see Supplementary Fig. S2). Notably, while a similar imbalance exists in the magnetic density data, it nonetheless exhibited a distribution that is considerably more clearly defined.

A few sample structures obtained from conditional generation are shown in Fig. 4d-f, with the generated charge density omitted from the visualization for clarity. Examination of the generated structures confirmed that they are chemically plausible and consistent with established chemical knowledge. For example, the structures generated to achieve high magnetic density targets included transition metals with partially filled 3d orbitals (e.g., $Fe_3O_4$), and those generated with high bandgaps showed strong ionic bonds with valence bands dominated by highly electronegative anions (e.g., $NaLiI_2$). Additionally, the structures generated with high target crystal density contained heavy elements, which favor close-packed metallic arrangements (e.g., HfCoPt).

A primary motivation for introducing charge density into the generation process was the expectation that explicitly capturing electronic structure would improve the model's inverse design capabilities. To investigate this, we conducted the ablation study to quantify the contribution of charge density in conditional generation performance. To date, generative models for inorganic materials have generally been compared only in terms of unconditional generation performance, and, to the best of our knowledge, systematic comparison of their conditional generation remain limited. Therefore, in this study, we defined a performance metric for evaluating conditional generation based on the proportion of generated structures whose property values fall within 10 % of the target, referred to as the success rate, to facilitate the comparison of inverse design capability between models.

Based on this metric, we conducted an ablation study on charge density to assess its impact on inverse design capability. The ablation study was performed for the same three properties (bandgap, magnetic density, and crystal density), with three different target values considered for each property (Fig. 4g). Across all three properties and target values, ChargeDIFF consistently achieved higher success rates than its ablated variant, ChargeDIFF (AXL), indicating a clear improvement in conditional generation performance. This gain is attributable to the direct relationship between a material's functionality and its electronic structure, with the incorporation of charge density enabling ChargeDIFF to more effectively capture electronic structure and thus achieve superior inverse design performance. In particular, for crystal density, although it is a structural property and might be presumed to have only a weak correlation with the electronic structure, the incorporation of charge density provides rich information on bond strength, coordination, and packing. Combined with the improved understanding of material symmetry observed during the latent diffusion process (Fig. 2b), this facilitates the generation of structures with the desired crystal density. Taken together, the pronounced improvements in conditional generation achieved by ChargeDIFF indicate that incorporating electronic structure-awareness can more effectively advance materials inverse design, which is the ultimate goal of generative modeling.

Furthermore, the conditioning performance of ChargeDIFF was compared with that of the previously reported MatterGen (Supplementary Fig. S12). ChargeDIFF showed performance broadly comparable to MatterGen, while exhibiting better conditioning performance on bandgap. This is plausibly due to the stronger correlation between bandgap and the electronic structure information provided by the incorporated charge density. It is particularly notable that MatterGen was pre-trained on a massive dataset of approximately 650,000 structures, whereas ChargeDIFF was trained on the much smaller MP-20-Charge dataset (40,516 structures), further highlighting its strong inverse design capability. Beyond this, ChargeDIFF also demonstrated robust performance in additional conditional generation tasks targeting energy above hull and space group (Supplementary Fig. S13 and S14), suggesting that inverse design can be extended to a broader set of target properties.

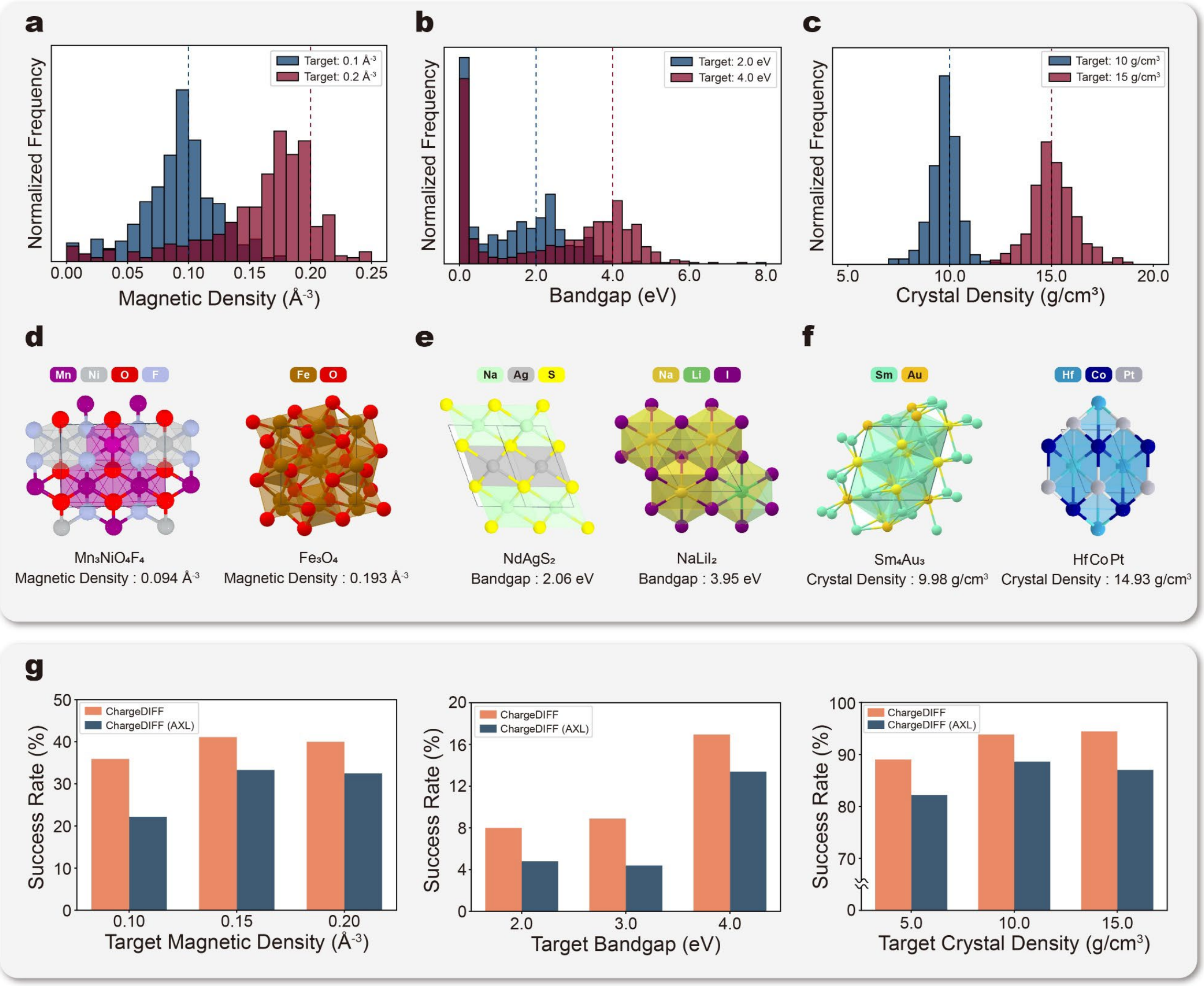

**Fig. 4. Conditional Generation Results. a-c.** Conditional generation results for magnetic density (**a**), bandgap (**b**), and crystal density (**c**). The property distributions of generated structures for different target values are shown. **d-f.** Example generated structures for magnetic density (**d**), bandgap (**e**), and crystal density (**f**) at each target value. For clarity, the generated charge density is omitted, and only the atomic structures are visualized. **g.** Ablation study on charge density. Conditional generation results for ChargeDIFF (orange) and its ablated variant without charge density, ChargeDIFF (AXL) (blue), are compared. Performance is measured by success rate, defined as the percentage of generated raw structures whose properties fall within 10 % of the target value.

**3D Charge Density-based Inverse Design of Functional Inorganic Materials**

The introduction of a new 3D voxel-based modality into the generation process enables new generative approaches for inorganic materials that were previously unattainable. In particular, we focus on an inpainting technique commonly used in image processing that fills in missing or masked regions in a given input, either based on surrounding context or guided by desired features[29,41]. By incorporating inpainting techniques into ChargeDIFF, it becomes possible to specify a target region within the unit cell and guide the charge density in that region toward a desired profile (e.g., low or high) during the generation process. We referred to this new materials generation approach as 3D charge density-based inverse design.

3D charge density-based inverse design can, in principle, be applied to a wide range of applications, but it is particularly well suited for fine-tuning transport behavior within a structure. In lithium-ion battery (LIB) cathode materials, for example, battery performance is strongly influenced by the dimensionality of the Li-ion migration pathways[42-45], and prior studies have reported that these pathways tend to follow regions of low charge density within the host material[46-48]. Based on these considerations, cathode materials with desired Li-ion migration pathways were designed using the 3D charge density-based inverse design approach (Fig. 5a). An inpainting technique was employed to specify a target region within the unit cell and suppress the charge density in that region toward zero during generation, which, in the final structure, can serve as Li-ion migration pathways. Additional conditioning on the chemical system was applied jointly during the generation process to ensure that the resulting structures are not composed of arbitrary atoms but instead belong to desired material classes—for example, transition metal oxides commonly used in practical batteries. Further details on the charge density-guided design process are provided in to Supplementary Information Section S5.

Using this approach, transition metal oxides with target ion migration channels of one-dimensional (1D) and two-dimensional (2D) were generated. For each case, 1,000 structures were generated, and the structural analysis of the Li coordination revealed that 323 (1D target) and 413 (2D target) structures possess Li-ion networks with the desired dimensionality. Additional nudged elastic band (NEB) calculations were performed to analyze the Li-ion migration energy barriers, identifying structures with energy barrier of below 0.5 eV (Fig. S17). Fig. 5b and 5c show representative examples of the generated structures and their Li-ion migration pathways obtained from molecular dynamics (MD) simulations. Notably, the obtained set includes novel structures that were not present in the training data, yet identical or closely similar structures have been investigated as LIB materials. For example, among the structures designed to exhibit 2D ion migration pathways, the layered lithium titanium oxide ($LiTi_2O_4$; Supplementary Fig. S18f) and layered lithium manganese oxide ($Li_3Mn_4O_8$; Supplementary Fig. S18g) possess structural analogues that have been experimentally examined as LIB cathodes[49,50], further underscoring the practical relevance of the proposed method.

Taken together, these results demonstrate that the 3D charge density-based inverse design strategy can be applied to generate functional materials by enabling precise tuning of their electronic structures. The approach is readily extensible to more intricate charge density targets (e.g., more complex migration pathways) and to additional material classes, including transition metal phosphates and polyanion compounds. To our knowledge, this represents the first implementation of inverse design targeting ion migration pathways. We anticipate that this approach will support the design of advanced battery materials while inspiring new generative modeling strategies of electronic materials in a broad range of applications.

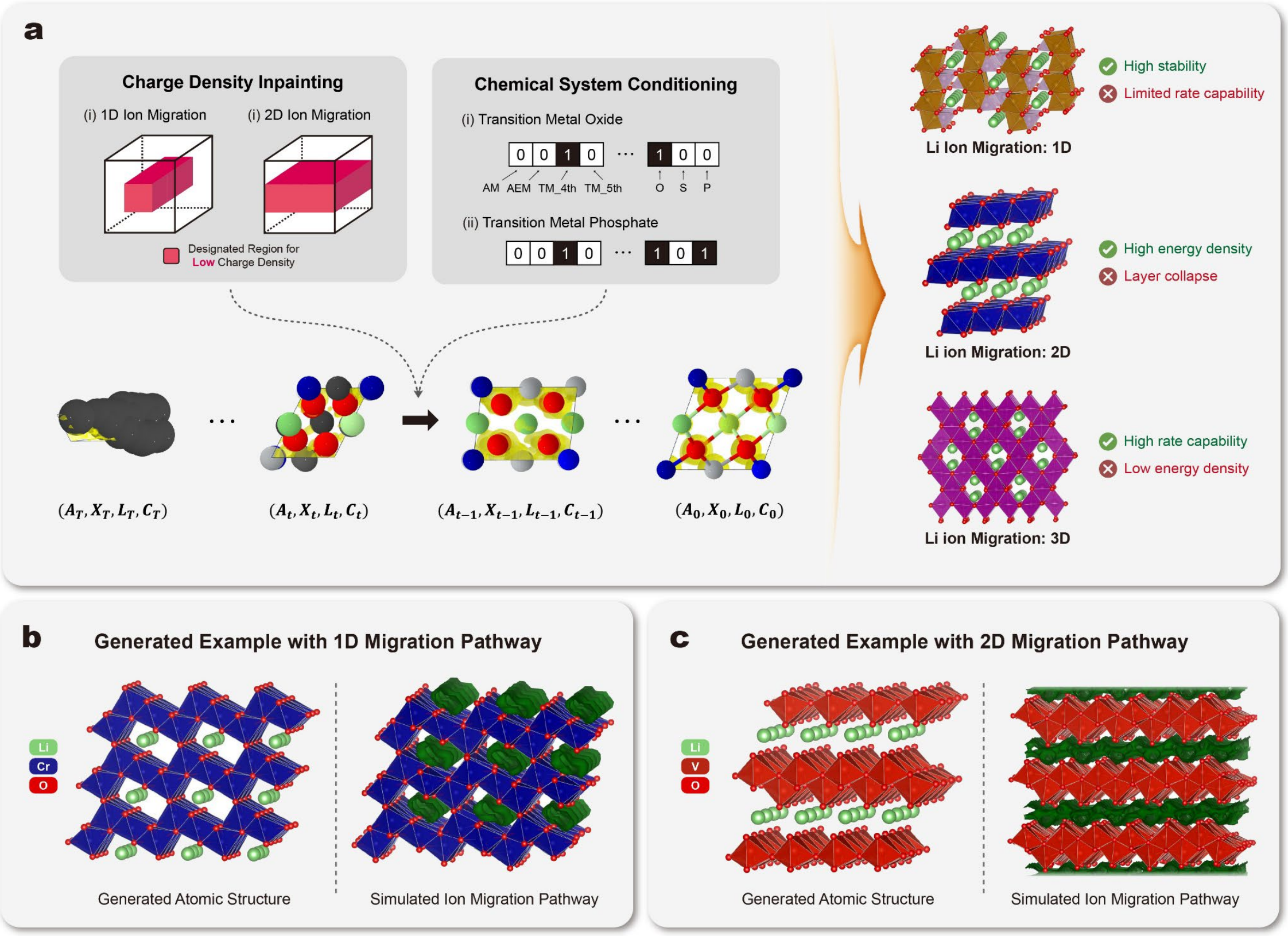

**Fig. 5. 3D Charge Density-based Inverse Design for Lithium-ion Battery Cathode Materials. a.** 3D charge density-based inverse design scheme. Using inpainting techniques, a target region within the unit cell is specified so that the charge density in that region is kept low during the generation process. The designated region serves as a lithium migration pathway, enabling the inverse design of cathode materials with a desired ion migration pathway topology. By additionally performing conditioning on the material's chemical system, the generated materials can be restricted to desired classes, such as transition metal oxides. **b-c.** Example cathode materials generated via 3D charge density-based inverse design targeting Li-ion migration pathway of 1D (b) and 2D (c) channels, respectively. Atomic structures are shown on the left (with charge density omitted for clarity), while the lithium-ion migration pathways obtained from molecular dynamics simulations are visualized in green on the right.

## Discussion

The generative design of inorganic materials holds immense potential to accelerate the discovery of new functional materials, driving innovation in industrial applications. ChargeDIFF introduces a new method in material representation for their generative modeling, which was previously limited to a structure-only representation, by explicitly incorporating charge density as an additional modality. Based on deeper understanding of the materials' electronic structure, ChargeDIFF exhibited exceptional generation performance, where this improvement is particularly pronounced in inverse design tasks targeting various chemical properties. Furthermore, the demonstration of charge density-based inverse design in LIB cathode materials highlights the broader potential of this methodology for a wider range of applications. For any applications where functionality is closely related to electronic structure—catalysis, energy materials, superconductors, semiconductors, and low dimensional systems—ChargeDIFF's capability of electronic structure-aware generative design could be impactful. The electronic-informed inverse design pipeline proposed in this work seamlessly bridges structural generation with functional electronic behavior.

Beyond charge density, incorporating various other types of materials properties as additional modalities in the generative process provides additional flexibility and potential. For instance, spin density[51] (representing the difference between up-spin and down-spin electrons) can be incorporated without any modification to the methodology, which is expected to provide even greater performance improvements when targeting properties related to the material's magnetism. Moreover, other types of data, including non-voxel representations, could also be incorporated into the generation process with minor modifications to the architecture, further broadening the model's capabilities for future direction.

From a broader perspective, a note can be made between generative and predictive models when incorporating additional modalities of data. For predictive models, additional inputs such as charge density may improve prediction performance, but they require complex feature preparation (e.g., DFT calculations) for every prediction, which is burdensome for general users. In contrast, for generative models, including such modalities may slightly increase the complexity of the training process, but imposes little to no additional cost during generation. (For instance, charge density is generated through denoising from random noise, without requiring DFT or other expensive computations.) This work shows that introducing additional modality of data such as charge density into generative model training is both effective and cost-efficient, making ChargeDIFF the first demonstration of such approach for future inverse materials discovery.

# Methods

## MP-20-Charge Dataset

In reference to the MP-20 dataset[21], which has been the most widely used benchmark for training generative models for inorganic materials, MP-20-Charge dataset was newly curated for training ChargeDIFF. Since the Materials Project[40] provides DFT-calculated charge density data for a substantial fraction of structures, we utilized this resource for training rather than recalculating charge densities. The MP-20-Charge dataset was constructed by querying the Materials Project database for structures that (1) contain 20 atoms or fewer, (2) have an energy above hull ≤ 0.1 eV/atom, (3) possess a longest cell length ≤ 20 Å, and, most importantly, (4) include available charge density data (as of March 21st, 2025). For charge density specifically, Fourier interpolation[52] was used to standardize the resolution to a uniform 32×32×32 grid, as detailed in the Supplementary Information Section S1. As a result, a total of 40,516 structures were collected to form the database. Given that the number of structures with available charge density data in the Materials Project continues to grow, the dataset size is expected to further expand in the future

## VQ-VAE

The VQ-VAE model was trained using charge density data within the MP-20-charge database, to map the original charge density resolution of 32×32×32 into latent resolution of 8×8×8. The dimensionality of the latent embedding space are treated as a hyperparameter; in this work, an embedding space of $e \in R^{8192\times3}$ was used. The total loss function is composed of a combination of reconstruction loss, vector quantization loss, and commitment loss as follows:

$$\mathbb{L} = \log p(x|z_q(x)) + \|sg[z_e(x)] - e\|_2^2 + \beta\|z_e(x) - sg[e]\|_2^2$$

where $z_e$ is the continuous encoder output, $z_q$ is the quantized latent obtained by replacing $z_e$ with its nearest codebook vector, and $e$ represents the learnable codebook embeddings that define the discrete latent space. Here, sg denotes the stop-gradient operation, and a $\beta$ value of 1.0 was used for training. An initial learning rate of $10^{-4}$ was set, with a scheduler that reduced the learning rate by a factor of 0.6 after 100 epochs of patience. Early stopping with a patience of 300 epochs was applied. A more detailed explanation of the VQ-VAE architecture can be found in its original paper[28].

## Diffusion Process of ChargeDIFF

In ChargeDIFF, each structure is represented by four components: atom type ($A \in \mathbb{A}^n$), atomic coordinate ($X \in [0,1)^{3\times n}$), lattice parameter ($L \in \mathbb{R}^{3\times 3}$), and charge density ($C \in \mathbb{R}^{32\times 32\times 32}$). Here, $M_t$ denotes the intermediate state of the material during the diffusion process at timestep $t$, which interpolates between the original material data ($M_0$) and prior distribution ($M_T$):

$$M_t = (A_t, X_t, L_t, C_t), \qquad 0 \leq t \leq T,$$

where T=1,000 was used for ChargeDIFF. The diffusion process in ChargeDIFF is defined by two Markov processes: the forward (noising) process, $q(M_t|M_{t-1})$, which is formulated independently for each component to gradually add noise to original data, and the backward (denoising) process, $p_\theta(M_{t-1}|M_t)$, iteratively removes the added noise to recover the original samples. The overall diffusion process is inspired by the formulation proposed in DiffCSP, with targeted modifications and improvements introduced in ChargeDIFF to accommodate its distinctive features, such as the inclusion of charge density. A more detailed formulation of the diffusion process for each component is provided in the subsequent sections.

### - Diffusion process of atom types (A)

Among the four components that constitute a material in ChargeDIFF, atom type is the only discrete component, and its diffusion process is modeled using D3PM[31], a discrete variant of DDPM[32], to account for this discrete nature. The forward process can be represented as:

$$q(A_t|A_0) = Cat(A_t; p = A_t\bar{Q}_t)$$

where a $Q_t$ is an absorbing transition matrix, $\bar{Q}_t = Q_1 Q_2 \dots Q_t$, and $Cat(x;p)$ is a categorical probability distribution with probabilities given by $p$. Specifically, the absorbing transition matrix is defined as:

$$[Q_t]_{ij} = \begin{cases} 1-\beta_t \ \ if \ \ i=j\neq 0 \\ 1 \ \ if \ \ i=j=0 \\ \beta_t \ \ if \ \ j=m, i\neq 0 \\ 0 \ \ else \end{cases}$$

where a cosine scheduler with the smoothing factor of 0.008 was used to schedule the $\beta_t$ values. This design ensures that $A_t$ gradually transitions to a dummy atom with an atomic number of 0 as $t$ becomes sufficiently large. To optimize the denoising process with regards to original distribution $q(A_0)$, a variational upper bound (VLB) on the negative log-likelihood is employed:

$$\mathbb{L}_{vb} = \mathbb{E}_{q(A_0)}\left[\begin{array}{c} D_{KL}[q(A_T|A_0)||p(A_T)] + \sum_{t=2}^{T} \mathbb{E}_{q(A_t|A_0)}[D_{KL}[q(A_{t-1}|A_t,A_0)||p_\theta(A_{t-1}|A_t)]] \\ -\mathbb{E}_{q(A_1|A_0)}[\log p_\theta(A_0|A_1)] \end{array}\right]$$

The total loss used for training is prepared by combining this VLB with an auxiliary cross-entropy loss:

$$\mathbb{L}_A = \mathbb{L}_{vb} + \lambda_{ce}\mathbb{E}_{q(A_0)}\mathbb{E}_{q(A_t|A_0)}[-\log p_\theta(A_0|A_t)]$$

where $\lambda_{ce}$ is the weight of the cross-entropy loss, set to 0.01 following common practice. Further details on D3PM algorithm can be found in the original paper.[31]

**- Diffusion process of atom types (X)**

Due to the periodicity of the crystal structure, the fractional coordinate of atoms is continuous variables bounded within $[0,1)^{3\times n}$, where n represents the number of atoms within the unit cell. Unlike other continuous parameters described later, the bounded nature of atomic coordinates motivates modeling their diffusion process using a Score-Matching (SM)[53] with a Wrapped Normal (WN) distribution[33]. The forward process of atomic coordinate is modeled as:

$$q(X_t|X_0) = \mathcal{N}_\omega(X_t|X_0, {\sigma_t}^2 I)$$

where $\mathcal{N}_\omega$ is a WN distribution with the values of standard deviation, $\sigma_t$, are assigned by an exponential scheduler within the range (0.005, 0.5). In practice, a sample from this distribution is efficiently obtained using a truncated wrapping function:

$$X_t = \omega(X_0 + \sigma_t \epsilon_X),\ \epsilon_X \sim \mathcal{N}(0, I)$$

where $\omega(\cdot)$ maps coordinates back into $[0,1)$. During the training phase, a denosing network predicts the score of the forward noising distribution, and the training objective is given by:

$$\mathbb{L}_X = \mathbb{E}_{q(X_0)} \mathbb{E}_{q(X_t|X_0)} \left[ \lambda_t \left\| \hat{\epsilon}_X(M_t, t) - \nabla_{X_t} \log q(X_t|X_0) \right\|_2^2 \right]$$

where $\lambda_t = \mathbb{E}^{-1}_{q(X_t|X_0)} \left[ \left\| \nabla_{X_t} \log q(X_t|X_0) \right\|_2^2 \right]$ is approximated through Monte-Carlo sampling.[22]

**- Diffusion process of atom types (L)**

The lattice parameter is an $\mathbb{R}^{3\times3}$ matrix encoding the unit cell geometry. Since the lattice parameter is a continuous variable, DDPM was applied for its diffusion process, which has been widely adopted in generative modeling of various classes of materials[54-56]. The forward process is progressively perturbs a clean lattice parameter $L_0$ towards the standard Gaussian prior $L_T \sim \mathcal{N}(0, \boldsymbol{I})$ as:

$$q(L_t|L_0) = \mathcal{N}\left(L_t; \sqrt{\bar{\alpha}_t} L_0, (1 - \bar{\alpha}_t) I\right)$$

where $\beta_t$ is a hyperparameter for the variance scheduling, and $\bar{\alpha}_t = \prod_{s=1}^{t} \alpha_s = \prod_{s=1}^{t} (1 - \beta_s)$. Here, $\beta_t$ values are assigned following a linear scheduler within the range of (0.0001, 0.02). Using the reparameterization trick[32], lattice parameter at any given timestep $t$ can be sampled as:

$$L_t = \sqrt{\bar{\alpha}} L_0 + \sqrt{1 - \bar{\alpha}_t} \epsilon_L\ , \qquad \epsilon_L \sim \mathcal{N}(0, I)$$

The backward process is defined as a parameterized Gaussian transition:

$$p_\theta(L_{t-1}|M_t) = \mathcal{N}(L_{t-1}; \mu_\theta(M_t), \sigma_t^2 \boldsymbol{I})$$

where $\mu_\theta(M_t) = \frac{1}{\sqrt{\alpha_t}}\left(M_t - \frac{\beta_t}{\sqrt{1-\bar{\alpha}_t}}\hat{\epsilon}_L(M_t, t)\right)$, and $\sigma_t^2 = \beta_t\frac{1-\bar{\alpha}_{t-1}}{1-\bar{\alpha}_t}$. $\hat{\epsilon}_L(M_t, t)$ denotes the noise predicted by the denoising model. The network is trained by minimizing the following mean squared error loss:

$$\mathbb{L}_L = \mathbb{E}_{q(L_0)}\mathbb{E}_{q(L_t|L_0)}(\|\hat{\epsilon}_L(M_t, t) - \epsilon_L\|_2^2)$$

**- Diffusion process of atom types (C)**

Although the charge density in the MP-20-Charge dataset is available at a resolution of 32×32×32, the diffusion process operates on a compressed 8×8×8 voxels obtained via pre-trained VQ-VAE encoder. This process occurs on the continuous latent embedding before quantization. Accordingly, similar to the lattice parameters, the diffusion process for charge density is formulated using a DDPM. The forward process is defined as:

$$q(C_t|C_0) = \mathcal{N}\left(C_t; \sqrt{\bar{\alpha}_t}C_0, (1-\bar{\alpha}_t)I\right)$$

where $\bar{\alpha}_t = \prod_{s=1}^{t}\alpha_s = \prod_{s=1}^{t}(1-\beta_s)$. Here, $\beta_t$ values are assigned following a linear scheduler ranging from 0.0001 to 0.02 over 1000 time steps, as in the case of lattice parameter diffusion. (Adopting a more customized noise schedule, similar to MatterGen[23], constitutes a promising avenue for future research.) The backward process is parameterized as:

$$p_\theta(C_{t-1}|M_t) = \mathcal{N}(C_{t-1}; \mu_\theta(M_t), \sigma_t^2\boldsymbol{I})$$

with $\mu_\theta(M_t) = \frac{1}{\sqrt{\alpha_t}}\left(M_t - \frac{\beta_t}{\sqrt{1-\bar{\alpha}_t}}\hat{\epsilon}_C(M_t, t)\right)$, and $\sigma_t^2 = \beta_t\frac{1-\bar{\alpha}_{t-1}}{1-\bar{\alpha}_t}$. Here, $\hat{\epsilon}_C(M_t, t)$ denotes the predicted noise by the denoising network, and training objective is given by:

$$\mathbb{L}_C = \mathbb{E}_{q(C_0)}\mathbb{E}_{q(C_t|C_0)}(\|\hat{\epsilon}_C(M_t, t) - \epsilon_C\|_2^2)$$

## Denoising Network

The denoising network is built based on the equivariant GNN (EGNN[57]), in which atoms and charge density probes are represented as distinct types of nodes. As explained in the main text, charge probe nodes act differently during the denoising process, with their coordinates remaining fixed at the grid points while only their features are updated. In contrast, atom nodes are updated both in terms of their coordinates and features during the denoising process. This architecture is inspired by previous works that predicted charge density of a given structure at specific spatial query locations[34,35]. To satisfy the periodic translational invariance defined in DiffCSP[22], the Fourier transformations are applied to encode inter-node distances during the message passing process. By leveraging the periodic nature of the trigonometric functions used in the expansion, the model ensures that periodic invariance is preserved throughout the generation process.

The overall training loss of the denoising network is defined as a combination of the losses previously defined for each component in the diffusion processes described in the preceding sections:

$$\mathbb{L}_{tot} = \omega_A \mathbb{L}_A + \omega_X \mathbb{L}_X + \omega_L \mathbb{L}_L + \omega_C \mathbb{L}_C$$

where $\omega_A = 10.0$, $\omega_X = 1.0$, $\omega_L = 1.0$, and $\omega_C = 1.0$ were used for the training. (Empirically, the model's performance was not highly sensitive to subtle changes in these parameters.) The network was trained with an initial learning rate of $5 \times 10^{-4}$, which was reduced by a factor of 0.6 using a scheduler with a patience of 100 epochs. Early stopping was applied with a patience of 300 epochs. A more detailed description of the denoising network architecture, along with several architectural features for more efficient generation, is provided Supplementary Information S2-2.

## Self-consistency Assessment of Charge Density

To verify the self-consistency between the generated charge densities and atomic structures, DFT calculations were performed to obtain the charge density for structures generated by ChargeDIFF, and the deviations from the generated charge densities were evaluated. The mean square deviation was defined as follows and used for the evaluation:

$$\varepsilon_{msd} = \frac{\int_{\vec{r}\in V}(c_{DFT}(\vec{r}) - c_{Generated}(\vec{r}))^2}{\int_{\vec{r}\in V}(c_{DFT}(\vec{r}))^2}$$

where $c_{Generated}$ is the charge density generated by ChargeDIFF, and $c_{DFT}$ is the charge density obtained from DFT. This metric was also used to analyze the results of the ablation study on the VQ-VAE.

## Computational Simulations

All DFT calculations were performed using the Vienna ab initio simulation package (VASP)[30] within the projector augmented-wave formalism[58], employing the automated workflow of Atomate2[59]. The calculation settings, including the use of the PBE functional[60] and the implementation of the Hubbard U correction, were kept consistent with the MP. For structure optimization process, a pre-relaxation was first performed using MatterSim[61], followed by the DFT relaxation using MPGGADoubleRelaxMaker workflow. For the static calculations used to obtain charge densities of the structures and various chemical properties, the MPGGAStaticMaker workflow was used.

To calculate the Li-ion migration barrier, the climbing image nudged elastic band (CI-NEB)[62] calculations were performed. Specifically, a single Li vacancy diffusion to a neighboring vacancy site was simulated, and the 10 nearest symmetrically unique hops for each material within a maximum distance of 4 Å were considered. The host crystal structure was expanded to a near-cubic supercell with a cell length of at least 10 Å to minimize interactions between periodic images. Seven intermediate images were used in the CI-NEB calculations, for which the image-dependent pair potential (IDPP)[63] method was employed for the initial guess.

MD simulations were conducted to obtain the explicit migration pathway for the selected materials. The simulations were performed for the crystal structure with one Li vacancy, and the NVT ensemble was used with a Nosé–Hoover thermostat[64,65]. Both CI-NEB calculations and MD simulations were performed using the atomic simulation environment (ASE)[66], employing MatterSim as a surrogate potential.

## Acknowledgements

This work was supported by National Research Foundation of Korea (RS-2024-00464386) funded by the Korea Government

## Data Availability

The datasets used for the training ChargeDIFF is available at https://figshare.com/ndownloader/files/58973161. Source Data are provided with this paper.

## Code Availability

A GitHub repository containing the source code and demos for the materials generation is available at https://github.com/parkjunkil/ChargeDIFF.

## Author information

**Corresponding Author**

Yousung Jung - *Department of Chemical and Biological Engineering, Seoul National University, 1 Gwanak-ro, Gwanak-gu, Seoul 08826, Korea; Institute of Engineering Research, Seoul National University, 1 Gwanak-ro, Gwanak-gu, Seoul 08826, Korea; Institute of Chemical Processes, Seoul National University, 1 Gwanak-ro, Gwanak-gu, Seoul 08826, Korea*

**Authors**

Junkil Park - *Department of Chemical and Biological Engineering, Seoul National University, 1 Gwanak-ro, Gwanak-gu, Seoul 08826, Korea; Institute of Engineering Research, Seoul National University, 1 Gwanak-ro, Gwanak-gu, Seoul 08826, Korea*

Junyoung Choi - *Department of Chemical and Biological Engineering, Seoul National University, 1 Gwanak-ro, Gwanak-gu, Seoul 08826, Korea*

**Contributions**

J.P. conceptualized the project, developed the code, and carried out the analysis. J.C. carried out the NEB calculations and MD simulations. Y.J. supervised the project. The manuscript was written with contributions from all authors.

# Ethics declaration

### Competing interests

The authors declare no competing interests.

# Reference

1 Zunger, A. Inverse design in search of materials with target functionalities. *Nature Reviews Chemistry* **2**, 0121 (2018).

2 Noh, J., Gu, G. H., Kim, S. & Jung, Y. Machine-enabled inverse design of inorganic solid materials: promises and challenges. *Chemical Science* **11**, 4871-4881 (2020).

3 Wang, J., Wang, Y. & Chen, Y. Inverse design of materials by machine learning. *Materials* **15**, 1811 (2022).

4 Balaraman, A. A. & Dutta, S. Inorganic dielectric materials for energy storage applications: a review. *Journal of Physics D: Applied Physics* **55**, 183002 (2022).

5 Murdock, B. E., Toghill, K. E. & Tapia-Ruiz, N. A perspective on the sustainability of cathode materials used in lithium-ion batteries. *Advanced Energy Materials* **11**, 2102028 (2021).

6 Xiang, Y. *et al.* Application of inorganic materials as heterogeneous cocatalyst in Fenton/Fenton-like processes for wastewater treatment. *Separation and Purification Technology* **295**, 121293 (2022).

7 Zhong, M. *et al.* Accelerated discovery of CO2 electrocatalysts using active machine learning. *Nature* **581**, 178-183 (2020).

8 Wang, S., Sun, M. & Hung, N. T. Vol. 12 81 (MDPI, 2024).

9 Kim, K. S. *et al.* The future of two-dimensional semiconductors beyond Moore's law. *Nature nanotechnology* **19**, 895-906 (2024).

10 Lamberti, F. *et al.* Design of experiment: a rational and still unexplored approach to inorganic materials' synthesis. *Sustainable Chemistry* **3**, 114-130 (2022).

11 Luo, S., Li, T., Wang, X., Faizan, M. & Zhang, L. High-throughput computational materials screening and discovery of optoelectronic semiconductors. *Wiley Interdisciplinary Reviews: Computational Molecular Science* **11**, e1489 (2021).

12 Ahmed, A. *et al.* Exceptional hydrogen storage achieved by screening nearly half a million metal-organic frameworks. *Nature communications* **10**, 1568 (2019).

13 Merchant, A. *et al.* Scaling deep learning for materials discovery. *Nature* **624**, 80-85 (2023).

14 Davies, D. W. *et al.* Computational screening of all stoichiometric inorganic materials. *Chem* **1**, 617-627 (2016).

15 Li, Z. *et al.* Materials generation in the era of artificial intelligence: A comprehensive survey. *arXiv preprint arXiv:2505.16379* (2025).

16 Park, H., Li, Z. & Walsh, A. Has generative artificial intelligence solved inverse materials design? *Matter* **7**, 2355-2367 (2024).

17 Noh, J. *et al.* Inverse design of solid-state materials via a continuous representation. *Matter* **1**, 1370-1384 (2019).

18 Ren, Z. *et al.* An invertible crystallographic representation for general inverse design of inorganic crystals with targeted properties. *Matter* **5**, 314-335 (2022).

19 Nouira, A., Sokolovska, N. & Crivello, J.-C. Crystalgan: learning to discover crystallographic structures with generative adversarial networks. *arXiv preprint arXiv:1810.11203* (2018).

20 Kim, S., Noh, J., Gu, G. H., Aspuru-Guzik, A. & Jung, Y. Generative adversarial networks for crystal structure prediction. *ACS central science* **6**, 1412-1420 (2020).

21 Xie, T., Fu, X., Ganea, O.-E., Barzilay, R. & Jaakkola, T. Crystal diffusion variational autoencoder for periodic material generation. *arXiv preprint arXiv:2110.06197* (2021).

22 Jiao, R. *et al.* Crystal structure prediction by joint equivariant diffusion. *Advances in Neural Information Processing Systems* **36**, 17464-17497 (2023).
23 Zeni, C. *et al.* A generative model for inorganic materials design. *Nature*, 1-3 (2025).
24 Yang, S. *et al.* Scalable diffusion for materials generation. *arXiv preprint arXiv:2311.09235* (2023).
25 Joshi, C. K. *et al.* All-atom Diffusion Transformers: Unified generative modelling of molecules and materials. *arXiv preprint arXiv:2503.03965* (2025).
26 Dong, R., Fu, N., Siriwardane, E. M. & Hu, J. Generative Design of inorganic compounds using deep diffusion language models. *The Journal of Physical Chemistry A* **128**, 5980-5989 (2024).
27 Park, H., Onwuli, A. & Walsh, A. Exploration of crystal chemical space using text-guided generative artificial intelligence. *Nature Communications* **16**, 4379 (2025).
28 Van Den Oord, A. & Vinyals, O. Neural discrete representation learning. *Advances in neural information processing systems* **30** (2017).
29 Avrahami, O., Lischinski, D. & Fried, O. in *Proceedings of the IEEE/CVF conference on computer vision and pattern recognition.* 18208-18218.
30 Kresse, G. & Furthmüller, J. Efficiency of ab-initio total energy calculations for metals and semiconductors using a plane-wave basis set. *Computational materials science* **6**, 15-50 (1996).
31 Austin, J., Johnson, D. D., Ho, J., Tarlow, D. & Van Den Berg, R. Structured denoising diffusion models in discrete state-spaces. *Advances in neural information processing systems* **34**, 17981-17993 (2021).
32 Ho, J., Jain, A. & Abbeel, P. Denoising diffusion probabilistic models. *Advances in neural information processing systems* **33**, 6840-6851 (2020).
33 Jing, B., Corso, G., Chang, J., Barzilay, R. & Jaakkola, T. Torsional diffusion for molecular conformer generation. *Advances in neural information processing systems* **35**, 24240-24253 (2022).
34 Jørgensen, P. B. & Bhowmik, A. DeepDFT: Neural message passing network for accurate charge density prediction. *arXiv preprint arXiv:2011.03346* (2020).
35 Koker, T., Quigley, K., Taw, E., Tibbetts, K. & Li, L. Higher-order equivariant neural networks for charge density prediction in materials. *npj Computational Materials* **10**, 161 (2024).
36 Rombach, R., Blattmann, A., Lorenz, D., Esser, P. & Ommer, B. in *Proceedings of the IEEE/CVF conference on computer vision and pattern recognition.* 10684-10695.
37 Podell, D. *et al.* Sdxl: Improving latent diffusion models for high-resolution image synthesis. *arXiv preprint arXiv:2307.01952* (2023).
38 Maaten, L. v. d. & Hinton, G. Visualizing data using t-SNE. *Journal of machine learning research* **9**, 2579-2605 (2008).
39 Gebauer, N., Gastegger, M. & Schütt, K. Symmetry-adapted generation of 3d point sets for the targeted discovery of molecules. *Advances in neural information processing systems* **32** (2019).
40 Jain, A. *et al.* Commentary: The Materials Project: A materials genome approach to accelerating materials innovation. *APL materials* **1** (2013).
41 Lugmayr, A. *et al.* in *Proceedings of the IEEE/CVF conference on computer vision and pattern recognition.* 11461-11471.
42 Chen, T., Li, M. & Bae, J. Recent advances in lithium iron phosphate battery technology: a comprehensive review. *Batteries* **10**, 424 (2024).
43 Hoang, K. & Johannes, M. Tailoring native defects in LiFePO4: insights from first-principles calculations. *Chemistry of Materials* **23**, 3003-3013 (2011).
44 Wang, L., Chen, B., Ma, J., Cui, G. & Chen, L. Reviving lithium cobalt oxide-based lithium secondary batteries-toward a higher energy density. *Chemical Society Reviews* **47**, 6505-6602 (2018).
45 Liu, T. *et al.* Correlation between manganese dissolution and dynamic phase stability in spinel-based lithium-ion battery. *Nature Communications* **10**, 4721 (2019).
46 Liu, Y., Jiang, X., Zhao, J. & Hu, M. Electronic charge density as a fast approach for predicting Li-ion migration pathways in superionic conductors with first-principles level precision. *Computational Materials Science* **192**, 110380 (2021).
47 Rong, Z., Kitchaev, D., Canepa, P., Huang, W. & Ceder, G. An efficient algorithm for finding the minimum energy path for cation migration in ionic materials. *The Journal of chemical physics* **145** (2016).
48 Shen, J.-X., Horton, M. & Persson, K. A. A charge-density-based general cation insertion algorithm for generating new Li-ion cathode materials. *npj Computational Materials* **6**, 161 (2020).
49 Kim, D. *et al.* 2D Metal-Oxide Nanosheets as a Homogeneous Li-Ion Flux Regulator for High-Performance Anodeless Lithium Metal Batteries. *ACS nano* **18**, 23277-23288 (2024).
50 Liu, S. *et al.* Reviving the lithium-manganese-based layered oxide cathodes for lithium-ion batteries. *Matter* **4**, 1511-1527 (2021).

51 Ruiz, E., Cirera, J. & Alvarez, S. Spin density distribution in transition metal complexes. *Coordination Chemistry Reviews* **249**, 2649-2660 (2005).
52 Shen, J.-X. *et al.* A representation-independent electronic charge density database for crystalline materials. *Scientific data* **9**, 661 (2022).
53 Song, Y. *et al.* Score-based generative modeling through stochastic differential equations. *arXiv preprint arXiv:2011.13456* (2020).
54 Hoogeboom, E., Satorras, V. G., Vignac, C. & Welling, M. in *International conference on machine learning.* 8867-8887 (PMLR).
55 Park, J., Gill, A. P. S., Moosavi, S. M. & Kim, J. Inverse design of porous materials: a diffusion model approach. *Journal of Materials Chemistry A* **12**, 6507-6514 (2024).
56 Park, J., Lee, Y. & Kim, J. Multi-modal conditional diffusion model using signed distance functions for metal-organic frameworks generation. *Nature Communications* **16**, 34 (2025).
57 Satorras, V. G., Hoogeboom, E. & Welling, M. in *International conference on machine learning.* 9323-9332 (PMLR).
58 Blöchl, P. E. Projector augmented-wave method. *Physical review B* **50**, 17953 (1994).
59 Ganose, A. *et al.* atomate2. (2024). https://doi.org:https://zenodo.org/records/10677081
60 Perdew, J. P., Burke, K. & Ernzerhof, M. Perdew, burke, and ernzerhof reply. *Physical Review Letters* **80**, 891 (1998).
61 Yang, H. *et al.* Mattersim: A deep learning atomistic model across elements, temperatures and pressures. *arXiv preprint arXiv:2405.04967* (2024).
62 Henkelman, G., Uberuaga, B. P. & Jónsson, H. A climbing image nudged elastic band method for finding saddle points and minimum energy paths. *The Journal of chemical physics* **113**, 9901-9904 (2000).
63 Smidstrup, S., Pedersen, A., Stokbro, K. & Jónsson, H. Improved initial guess for minimum energy path calculations. *The Journal of chemical physics* **140** (2014).
64 Nosé, S. A unified formulation of the constant temperature molecular dynamics methods. *The Journal of chemical physics* **81**, 511-519 (1984).
65 Hoover, W. G. Canonical dynamics: Equilibrium phase-space distributions. *Physical review A* **31**, 1695 (1985).
66 Larsen, A. H. *et al.* The atomic simulation environment—a Python library for working with atoms. *Journal of Physics: Condensed Matter* **29**, 273002 (2017).

# Integrating electronic structure into generative modeling of inorganic materials

*Junkil Park*[1,2], *Junyoung Choi*[1] *and Yousung Jung*[1,2,3*]

[1] Department of Chemical and Biological Engineering, Seoul National University, 1 Gwanak-ro, Gwanak-gu, Seoul 08826, Korea

[2] Institute of Engineering Research, Seoul National University, 1 Gwanak-ro, Gwanka-gu, Seoul 08826, Korea

[3] Institute of Chemical Processes, Seoul National University, 1 Gwanak-ro, Gwanka-gu, Seoul 08826, Korea

**Electronic Supplementary Information**

# Section S1. Details on Dataset Construction

## Section S1-1. MP-20-Charge Dataset

As described in the Methods section of the original manuscript, the MP-20-Charge dataset was newly curated for training ChargeDIFF by querying the Materials Project database for structures that (1) contain 20 atoms or fewer, (2) exhibit an energy above hull ≤ 0.1 eV/atom, (3) have the longest cell length ≤ 20 Å, and, most importantly, (4) include available charge density data (as of March 21$^{st}$, 2025). As a result, a total of 40,516 structures were collected to form the database.

Owing to the distinct characteristics of each material and the varying k-point sampling used in the DFT calculations for charge density, the charge densities available in the Materials Project exhibit different voxel resolutions across structures. To ensure consistent input dimensionality to the machine learning model, the charge density of each structure was standardized to a uniform grid of 32×32×32 prior to inclusion in the MP-20-Charge dataset. This standardization was performed using Fourier interpolation implemented in the pyrho library (Fig. S1).

A relatively modest resolution of 32×32×32 was chosen to mitigate the computational cost of the machine learning model. Interpolation to lower resolutions (e.g., 8×8×8) resulted in a noticeable loss of detailed charge density information, which can be qualitatively verified in Fig. S1, and is discussed further in Section S3-3. This trade-off between voxel resolution and computational cost motivated the adoption of a latent diffusion scheme.

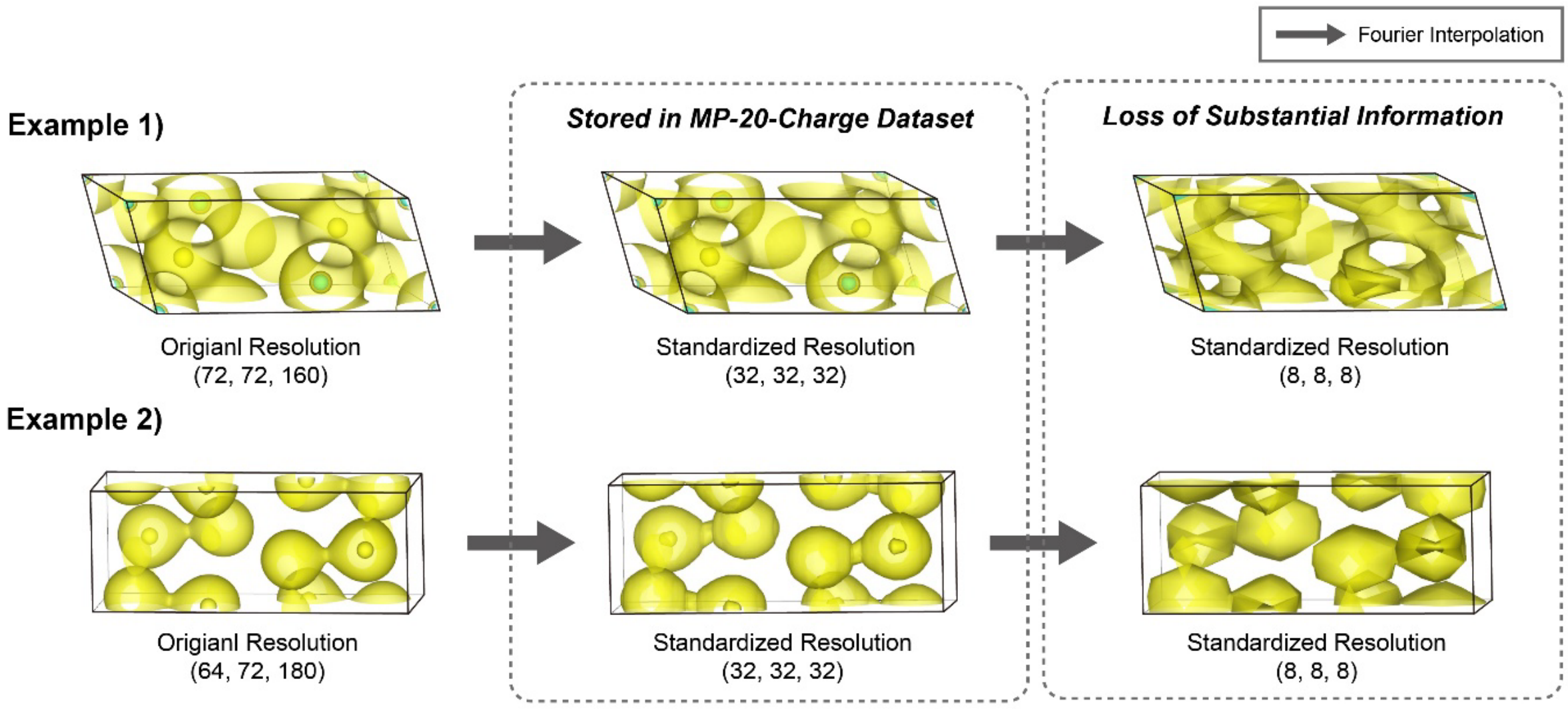

**Fig. S1. Charge density standardization for the MP-20-Charge dataset.** Charge density data from the Materials Project are standardized to a uniform 32×32×32 resolution using Fourier interpolation prior to their inclusion in the MP-20-Charge dataset. Interpolation to lower resolutions (e.g., 8×8×8) results in a substantial loss of fine charge density details, which can be visually observed as a degradation in the charge density profiles, similar to a loss of image resolution.

## Section S1-2. Characterization of MP-20-Charge Dataset

For the structures in the MP-20-Charge dataset, the distributions of magnetic density, bandgap, crystal density, energy above hull, and space group are shown in Fig. S2. These labeled data are subsequently used for training conditional generation models. Values for magnetic density, bandgap and energy above hull were obtained from the Materials Project, whereas crystal density and space group were computed using the pymatgen library.

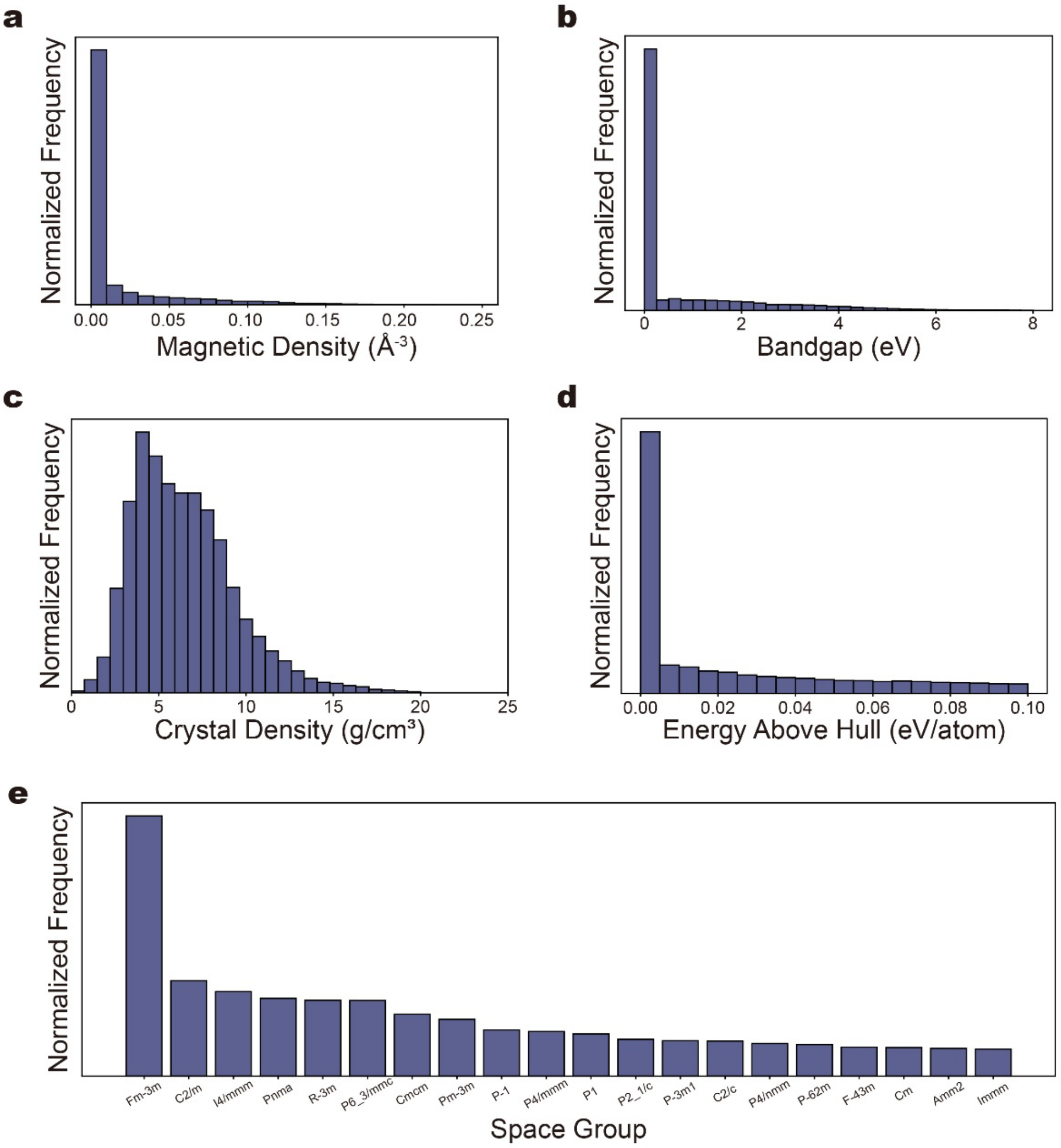

**Fig. S2. Characterization of the MP-20-Charge dataset. a-e.** Distributions of magnetic density (**a**), bandgap (**b**), crystal density (**c**), energy above hull (**d**), and space group (**e**) for the structures in the MP-20-Charge dataset. For the space group distribution, only the 20 most frequently occurring space groups are visualized.

# Section S2. Details on ChargeDIFF Architecture

## Section S2-1. Details on VQ-VAE

As described in Fig. 2 of the main text, the vector quantized-variational autoencoder (VQ-VAE) plays an essential role within the ChargeDIFF architecture by effectively compressing the charge density voxels while minimizing the information loss. Using VQ-VAE, the charge density at a resolution of 32×32×32 is mapped to a compressed latent space of 8×8×8, where the diffusion process is performed following a latent diffusion scheme similar to those used in high-resolution image generation processes (Fig S3). This latent diffusion setup reduces the number of charge density probes from 32×32×32 to 8×8×8, which brings the significant improvement in the efficiency of the training and inference phases.

The training of the VQ-VAE is conducted separately prior to the training of the ChargeDIFF denoising network, and the details of its training procedure are described in the Methods section of the main text. A qualitative analysis of the trained VQ-VAE was performed by comparing the visual attributes of reconstructed charge density voxels (i.e., sequential compression and restoration via encoder and decoder) with the original counterparts. As shown in Fig. S4, the reconstructed voxels exhibit nearly identical charge density distributions, which are barely distinguishable from the originals, providing direct evidence of the reconstruction performance of the trained VQ-VAE.

Additionally, similar to the latent space visualization by space group shown in Fig. 2b of the main text, Fig. S5 presents a t-SNE visualization illustrating how charge densities of the structures are mapped in the latent space according to their metallicity. In this process, the metallicity was roughly determined, with structures having a bandgap of 0 classified as metallic and those with a bandgap greater than 0 classified as non-metallic. While Fig. 2b highlights that the latent space captures structural features such as space group, Fig. S5 demonstrates that chemical properties, including metallicity, are also taken into account in the VQ-VAE latent representation.

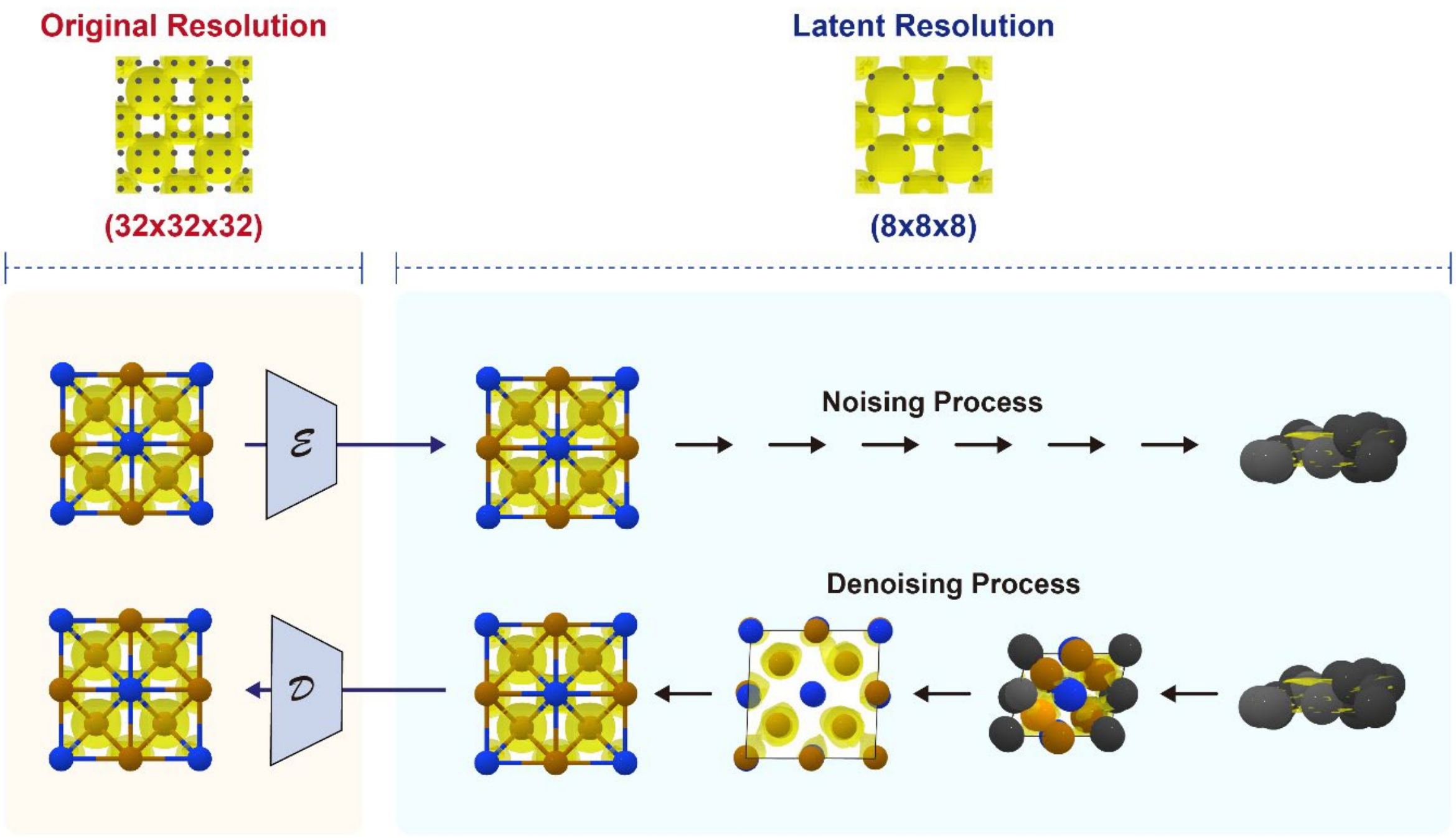

**Fig. S3. Latent diffusion setup within ChargeDIFF.** In ChargeDIFF, the diffusion process of charge density follows a latent diffusion scheme enabled by a trained VQ-VAE. The charge density data, originally at a resolution of 32×32×32, is mapped to an 8×8×8 space through the VQ-VAE encoder, and the diffusion process (including both noising and denoising process) is performed in this compressed latent space. After the generation process is complete, the charge density is restored to its original resolution through the decoder.

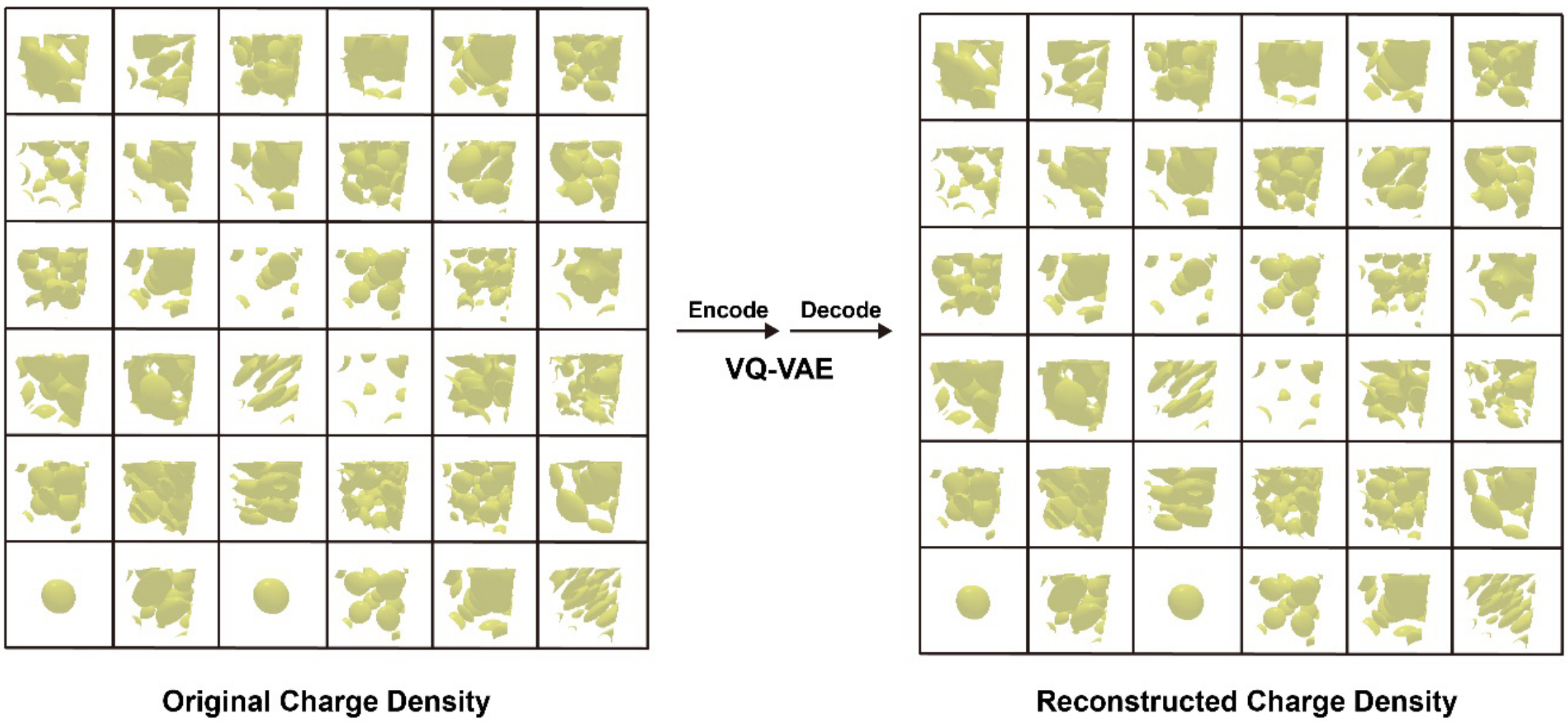

**Fig. S4. Reconstruction of Charge Density via VQ-VAE.** Qualitative comparison between the original charge density and its reconstruction through the trained VQ-VAE. An isosurface at 60 e/Å$^3$ is visualized for both. The reconstructed charge density closely matches the original, indicating that the VQ-VAE effectively compresses and restores the data with minimal information loss.

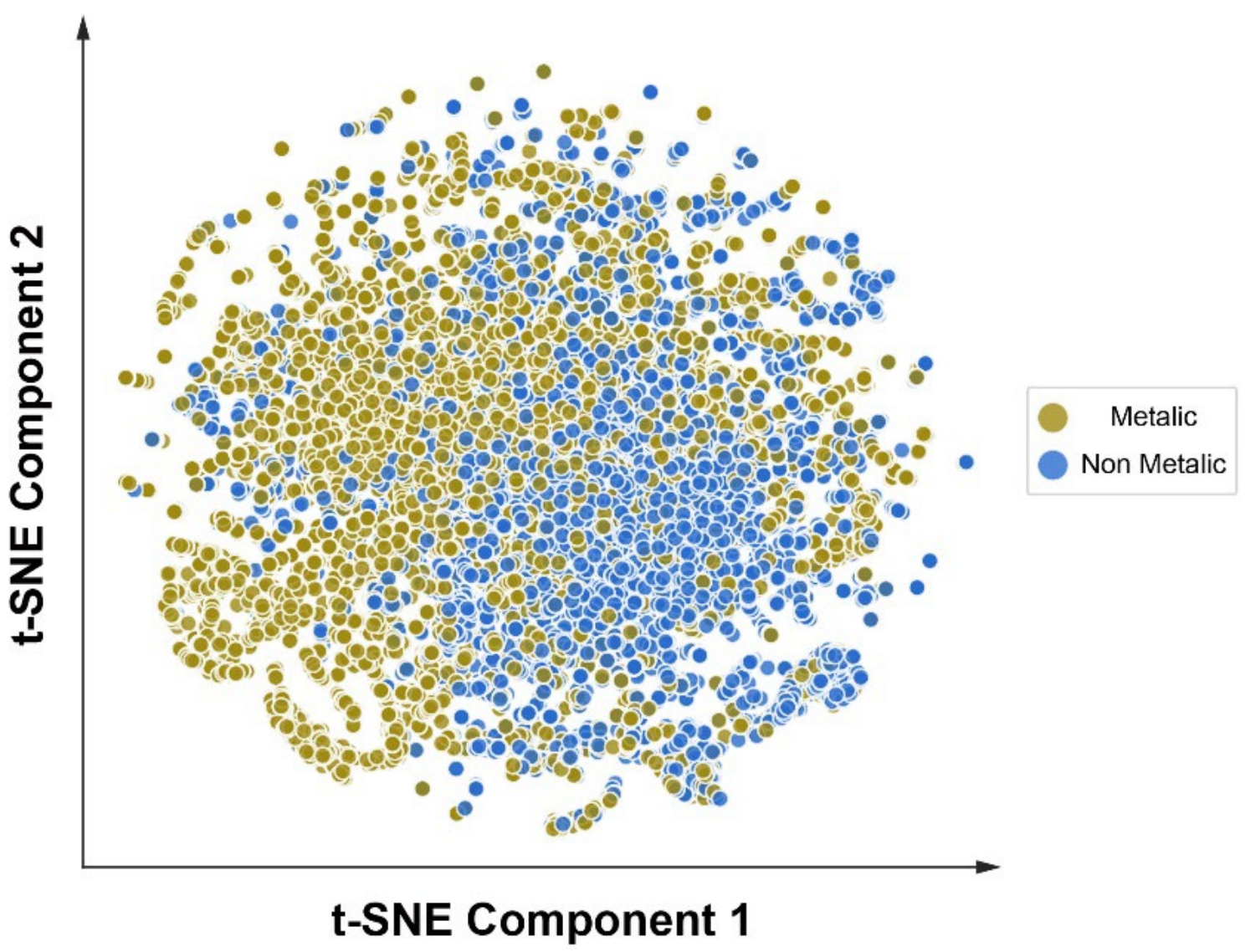

**Fig. S5. t-SNE visualization of VQ-VAE latent space by material metallicity.** Latent vectors of charge density within the training dataset encoded by the VQ-VAE were visualized by t-SNE, with materials classified according to their metallicity. Metallicity was roughly determined from the MP database bandgap values: materials with a band gap of 0 were classified as metallic, and those with a bandgap greater than 0 as non-metallic.

## Section S2-2. Details on Denoising Network

Training a diffusion model is equivalent to learning a denoising network that predicts the structure at the previous timestep $(t-1)$ by removing one step of noise from the structure at the current timestep $(t)$. The formulation of the diffusion process for each material component is described in the Methods section of the main text. As described in the main text, the denoising network of ChargeDIFF takes the form of an equivariant graph neural network, treating each atom and charge density probe as nodes with distinct modalities. Through their interactions, specifically cross-modal message passing (Fig. 1c), the structural and electronic components mutually influence each other during the generation process, enabling the generation of structures whose electronic structures are informed during the generation.

Looking more closely at the denoising network architecture, the material graph at any given timestep consists of up to 20 atom nodes (limited by the MP-20-Charge dataset) and a fixed number of 512 probe nodes (Fig. S6a). Atom-atom interactions are fully connected, while atom-probe interactions are restricted to the 20 nearest neighbors for each atom, as these local interactions are the most significant and it help maintain manageable computational cost. One noteworthy point is that the positions of the probe nodes remain fixed during the generation process. This allows the nearest neighbor search between atoms and probes to efficiently identify each atom's closest probe nodes without computing distances between all atom-probe pairs. After message passing, atom and probe nodes are processed through the atom model and probe model, respectively, and this whole message passing process is repeated three times. Finally, the updated node features are passed through the decoder to read out each material component (A, X, L, and C).

A part omitted in the previous explanation is the self-interaction between probe nodes (probe-probe interactions). To efficiently handle interactions among a large number of probe nodes, ChargeDIFF leverages their voxel-based nature. By taking advantage of the fixed coordinates of probe nodes, the nearest-neighbor message passing process is replaced by reshaping the probe features into their original voxel form and passing them through consecutive convolution layers, after which the features are flattened back (Fig. S6b). This convolutional processing is, in essence, equivalent to message passing between nearest neighbors, with circular padding used to enforce periodic boundary conditions. This approach bypasses the usual neighbor search, allowing feature updates to occur much faster and making the overall denoising process significantly more efficient.

The denoising network architecture of ChargeDIFF is designed to achieve the periodic translation invariance of the generated structures. Despite the inclusion of the additional charge density modality, invariance is easily maintained by treating charge density probes as distinct nodes in the graph, and applying the same procedures used for atom nodes. For example, in DiffCSP, when considering distances between nodes during message passing, a Fourier transformation is applied to the relative fractional coordinates, enabling message passing that respects periodic boundary conditions. This method is consistently applied to both probe-atom and probe-probe interactions, ensuring that periodic translational invariance is preserved throughout the generation process.

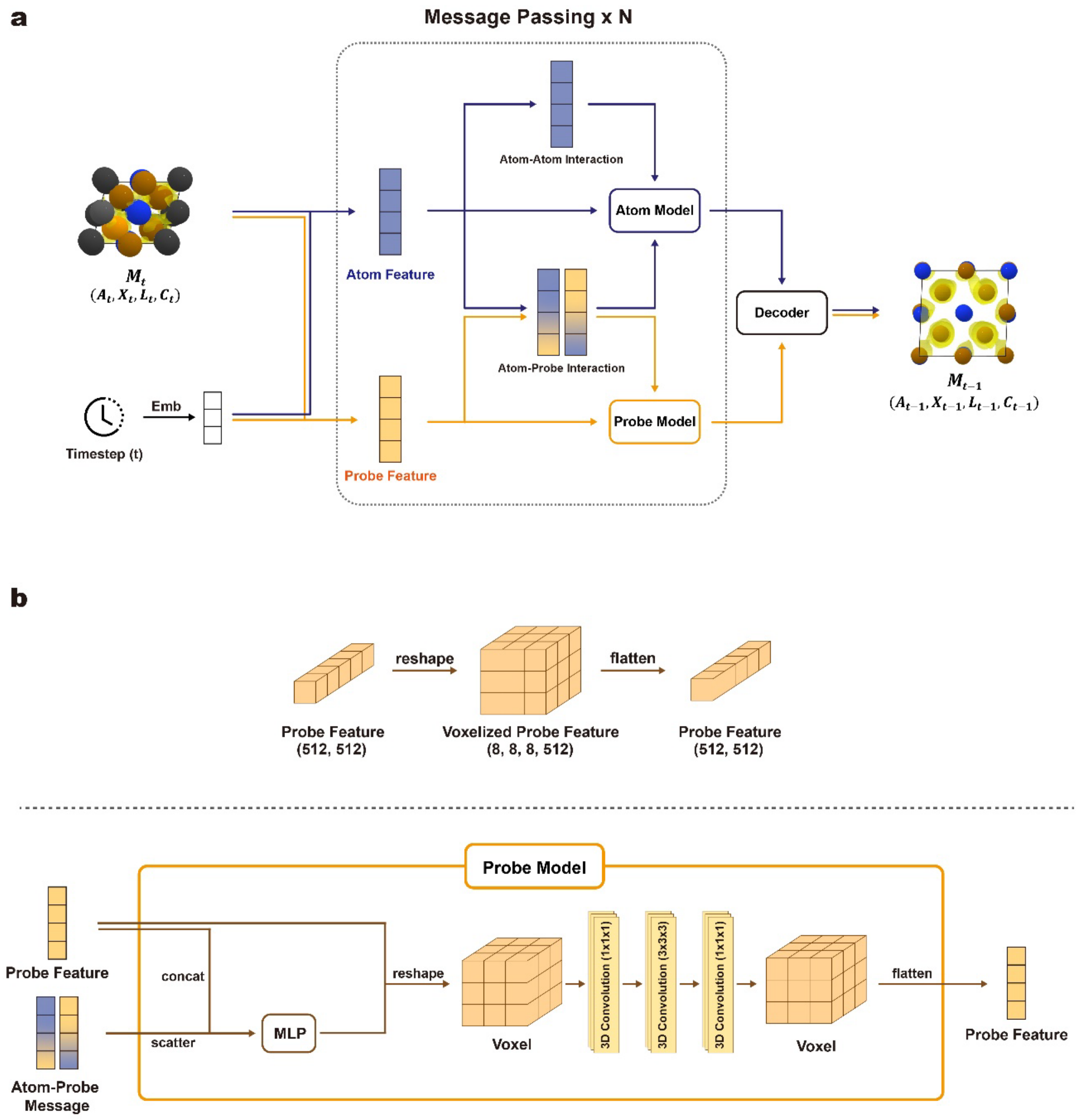

**Fig. S6. Denoising network of ChargeDIFF. a.** Detailed schematic of the denoising network. Atom features and charge density probe features dynamically influence each other through cross-modal interactions, enabling the network to predict the structure at the previous timestep with one step of noise removed. **b.** Voxelization of probe features within the probe model. Probe-probe interactions are replaced with an efficient convolution-based operation that leverages their voxel-based nature. This approach bypasses the distance-based neighbor search between probe nodes during the message passing, significantly improving the efficiency of the denoising procedure.

# Section S3. Details on Unconditional Generation

## Section S3-1. Characterization of the Generated Structures

The elemental distribution of 1,000 structures generated unconditionally using ChargeDIFF was visualized alongside that of structures in the MP-20-Charge dataset (Fig. S7). For the MP-20-Charge dataset, a broad elemental distribution across the periodic table was observed. Similarly, since ChargeDIFF was trained on this dataset, the generated structures were found to contain a wide variety of elements across the periodic table, demonstrating ChargeDIFF's capability of generating chemically diverse inorganic materials.

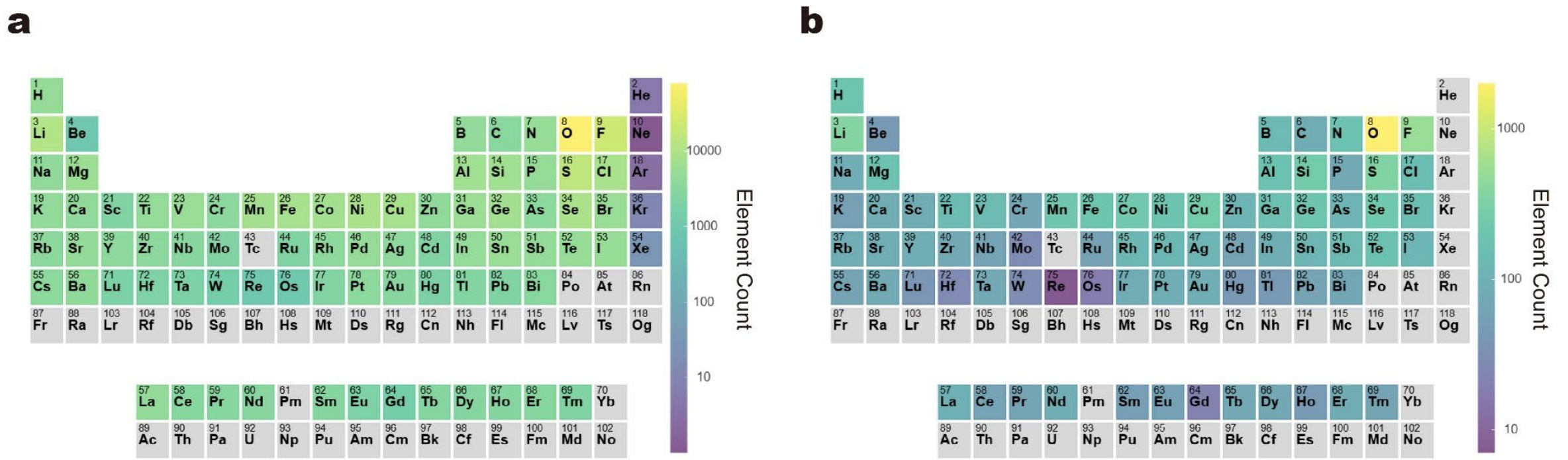

**Fig. S7. Elemental distribution of the generated structures. a-b.** Distribution of structures in the MP-20-Charge dataset **(a)**, and of 1000 structures generated unconditionally using ChargeDIFF **(b)**.

## Section S3-2. Details on Evaluation Metrics

To assess the model performance in unconditional generation and compare it with baseline models, the percentage of stable, unique, and new (SUN) materials among the generated structures were computed. The SUN metric is currently the most widely used metric for evaluating inorganic materials generative models. A structure is considered stable if its energy after DFT relaxation lies within 0.1 eV/atom above the convex hull. A structure is considered unique if it does not duplicate any other generated structures generated by the same method. A structure is considered new if it does not match any structures in the reference dataset. The Alex-MP-ICSD dataset was used as a reference for defining the convex hull and determining the novelty of the structures, as it is, to the best of our knowledge, the largest inorganic materials dataset . Caution should be exercised when comparing the model performance, as it depends on the choice of the reference dataset. Notably, the evaluation metric used in this work is identical to that in MatterGen, and the baseline SUN ratios reported in the MatterGen paper were used for comparison.

## Section S3-3. Ablation Study on VQ-VAE

To evaluate the importance of the charge density compression via VQ-VAE in the overall ChargeDIFF workflow, an ablation study on the VQ-VAE architecture was conducted. In contrast to the original ChargeDIFF, where 32×32×32 charge density voxels are compressed to 8×8×8 via the VQ-VAE encoder, the ablated mode directly uses 8×8×8 voxels prepared by Fourier interpolation for training (Fig. S8a). This assesses the impact of information compression via the VQ-VAE through higher-resolution voxels (32×32×32) on the improvement of generation performance.

The performance of the models (ChargeDIFF, ChargeDIFF (w/o VQ-VAE)) were compared based on the mean square deviation from the DFT-calculated charge densities and SUN structure ratio (Fig. S8b). The mean square deviation reflects the self-consistency between the generated structures and charge densities, and is used to evaluate how accurately the model reproduces DFT charge densities. For the 1,000 generated structures, as mentioned in the main text, ChargeDIFF exhibited a mean square deviation of 2.2 %, whereas ChargeDIFF (w/o VQ-VAE) showed a deviation of 20.6 %. This demonstrates that the introduction of the VQ-VAE allows the model to utilize higher-resolution charge density information, resulting in more accurate and precise charge density. Furthermore, regarding the SUN structure ratio, ChargeDIFF achieved 24.8 %, whereas ChargeDIFF (w/o VQ-VAE) reached only 19.6 %, indicating a significantly lower proportion. These results highlight the importance of efficiently leveraging charge density information and underscore that the VQ-VAE is a key component contributing to ChargeDIFF's superior generation capabilities.

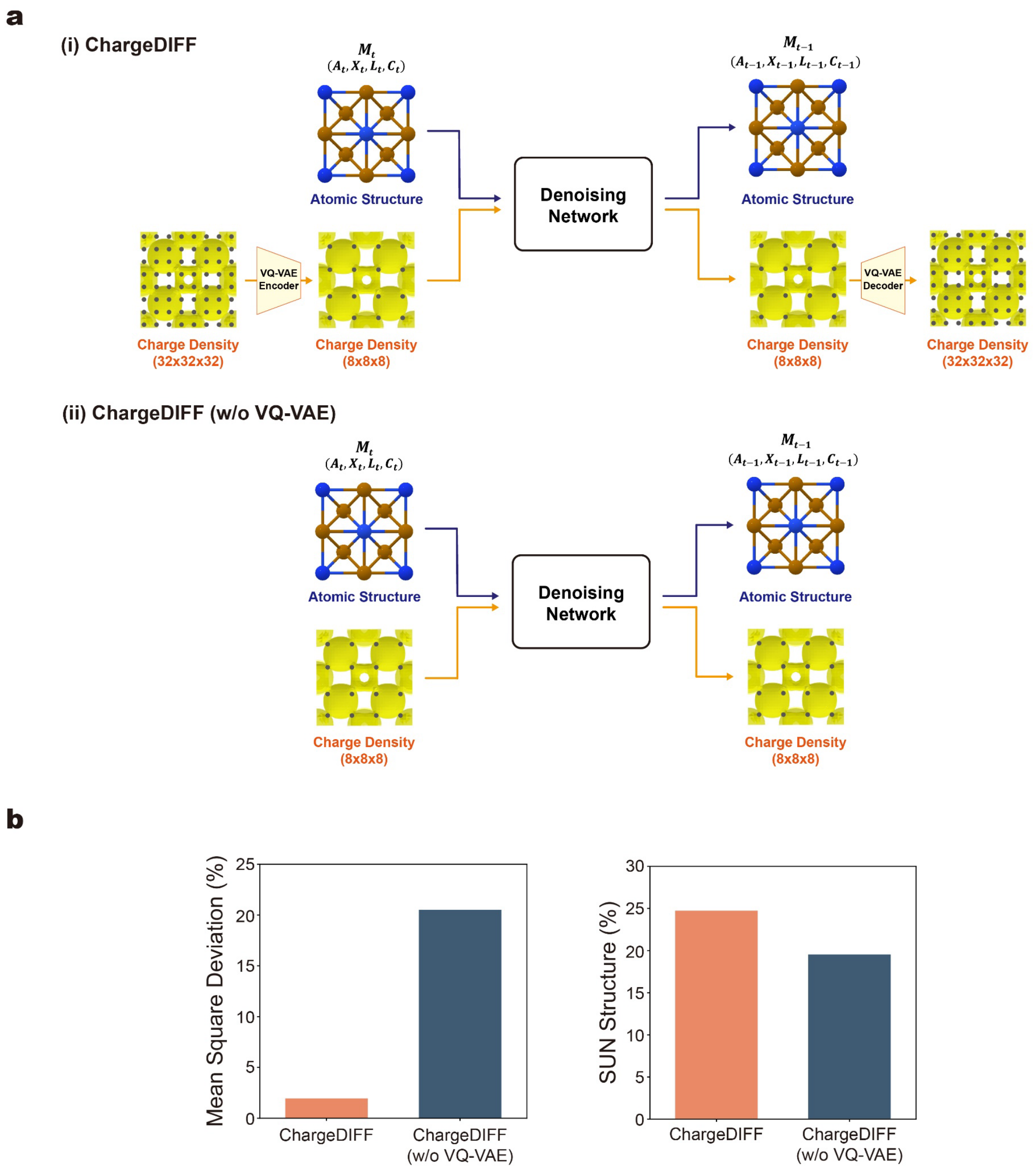

**Fig. S8. Ablation study on the VQ-VAE. a.** Denoising network architecture of ChargeDIFF and its ablated variant without the VQ-VAE (denoted as ChargeDIFF (w/o VQ-VAE)). In the ablated model, charge densities were directly prepared at an 8×8×8 resolution via interpolation and used for training. **b.** Performance comparison of ChargeDIFF (orange) and ChargeDIFF (w/o VQ-VAE) (blue) based on the mean square deviation from DFT-calculated charge densities (lower is better) and the SUN structure ratio among the generated structures (higher is better).

## Section S3-4. Inference Time Comparison

We compared the inference time for structure generation across ChargeDIFF, ChargeDIFF (AXL), and MatterGen. Specifically, we measured the time required to generate a single batch of size 100 using an NVIDIA GeForce RTX 4090 GPU. Due to the introduction of charge density, ChargeDIFF incurs a modest increase in inference time relative to ChargeDIFF (AXL); however, it achieves an approximately 56% reduction in inference time compared with MatterGen (Fig. S9). This indicates that, despite processing high-resolution charge density, ChargeDIFF enables efficient structure generation owing to architectural advantages, including latent diffusion.

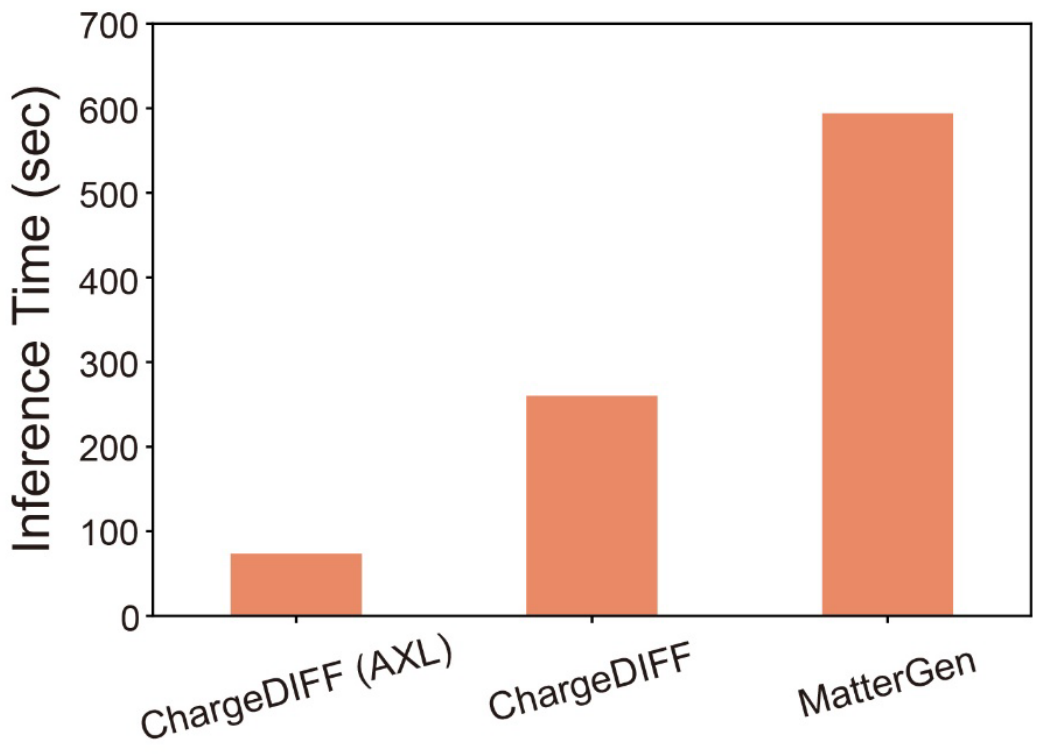

**Fig. S9. Inference time comparison for structure generation.** The wall-clock time to generate a single batch (n = 100) was measured for ChargeDIFF, ChargeDIFF (AXL; ablated without charge density), and MatterGen using an NVIDIA GeForce RTX 4090 GPU.

# Section S4. Details on Conditional Generation

## Section S4-1. Conditional Generation Model Architecture

The conditional ChargeDIFF model retains the original model architecture, while embedding information about the property of interest and concatenating it with the timestep embedding during training. During material generation, the desired target property is likewise embedded and provided as an additional input to the denoising network. The model is designed to accommodate target properties across multiple modalities, with modality-specific embedding schemes applied (Fig. S10). For example, a numeric property represented by real values (e.g., bandgap) is embedded using sinusoidal embeddings, whereas a categorical property (e.g., space group) is embedded through one-hot encoding followed by a linear layer. Conditioning on the chemical system (analogous to the crystal structure prediction task) is performed using a multi-hot encoding of the constituent elements. Although classifier-free guidance (CFG) has recently become a prevalent conditioning method, our experiments showed that it did not provide performance improvements for ChargeDIFF, and thus, it was not considered in this study.

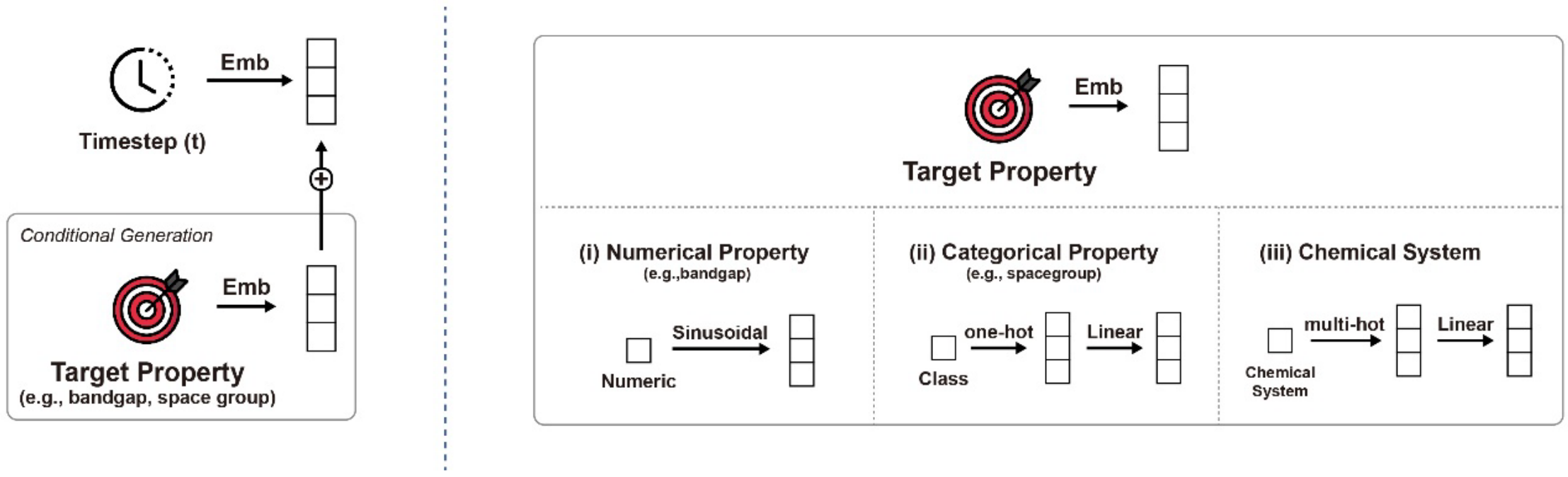

**Fig. S10. Conditioning for Diverse Data Modalities.** The inverse design in ChargeDIFF is achieved through conditional generation across multiple modalities. The conditional model does not require any changes to the underlying architecture. Instead, the target property is embedded and concatenated with the timestep embedding vector during training. Properties of different modalities are embedded according to methods suited to their respective characteristics.

## Section S4-2. Details on Evaluation Metrics

To evaluate the conditional generation performance of the models in Fig. 4a-c of the main text, 500 structures were generated for each target property, and the generated structures were relaxed using MatterSim. Magnetic density and bandgap were evaluated via DFT calculations, whereas crystal density was computed using the pymatgen library. For the detailed information on DFT calculations setup and structure relaxation process, please refer Methods section in the main text.

To compare conditioning performance across models, we defined a generation as successful if the property deviated by less than 10% from the target, and compared success rate among the generated structures. To assess pure conditioning performance without post-processing effects, the analysis in Fig. 4g was conducted on raw, unrelaxed structures. We repeated the same analysis after structure relaxation with MatterSim, and observed nearly identical trends (Fig. S11).

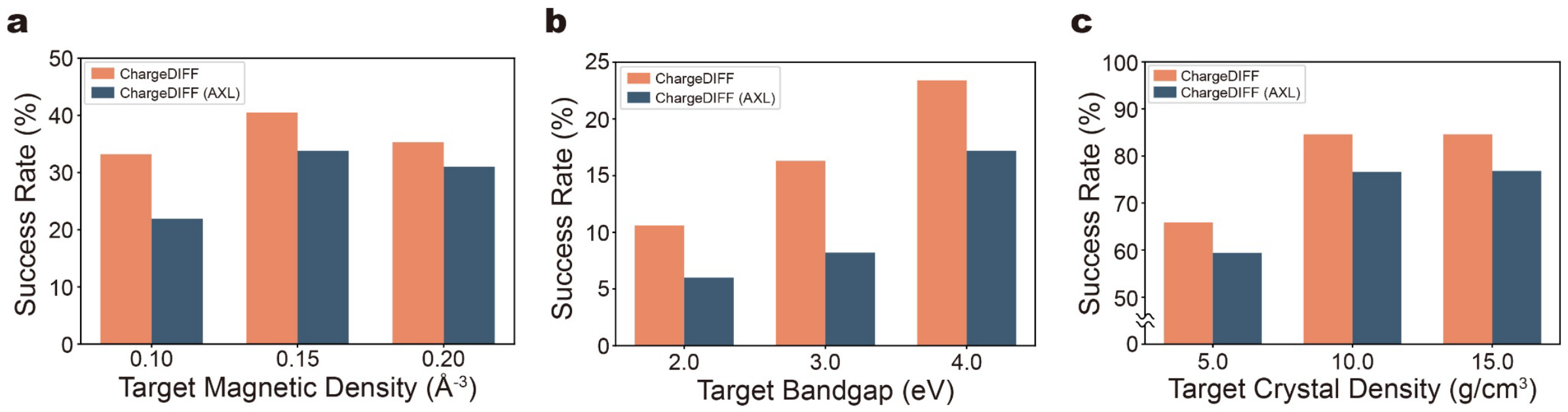

**Fig. S11. Conditional Generation Results after Structure Relaxation.** Conditional generation results for ChargeDIFF (orange) and its ablated variant without charge density, ChargeDIFF (AXL) (blue), are shown for magnetic density, bandgap, and crystal density. Performance was evaluated by success rate, defined as the percentage of generated structures whose properties fall within 10 % of the target value. Notably, the generated structures were relaxed using MatterSim, resulting in differences from those shown in Fig. 4g, where unrelaxed structures were evaluated.

## Section S4-3. Comparison with a Baseline

As noted in the main text, it is rare to compare the conditional generation performance of generative models, and, to our knowledge, no such comparisons have been reported for inorganic materials. Through a rigorous ablation study on charge density, we demonstrated the superior conditioning capability of ChargeDIFF and it relies critically on the incorporation of charge density. Building upon it, we compared the conditional generation performance of ChargeDIFF with MatterGen, which can be considered the strongest baseline.

Using the publicly available MatterGen model checkpoints trained on magnetic density and bandgap, we performed a performance comparison targeting these two properties. In their original work, the conditional MatterGen model is prepared slightly differently from ChargeDIFF. They first train an unconditional model on a large Alex-MP-20 database (607,634 structures) and then fine-tune it for conditional generation using the labeled data. The number of labeled data employed for fine-tuning varies between cases, with 605,000 labeled data were used for magnetic density, and 42,000 for bandgap.

The model performance was compared in terms of success rate, as in Fig 4g, and the results are shown in Fig. S12. As can be seen, ChargeDIFF showed performance broadly comparable to MatterGen, whereas for bandgap, ChargeDIFF outperformed MatterGen to some extent. These results are particularly notable given that the MatterGen model was pre-trained on a much larger dataset, and in the case of magnetic density, also fine-tuned with considerably more labeled data. This re-emphasize the outstanding inverse design capability of ChargeDIFF and indicates that training on more extensive database in the future could further improve its conditional generation performance.

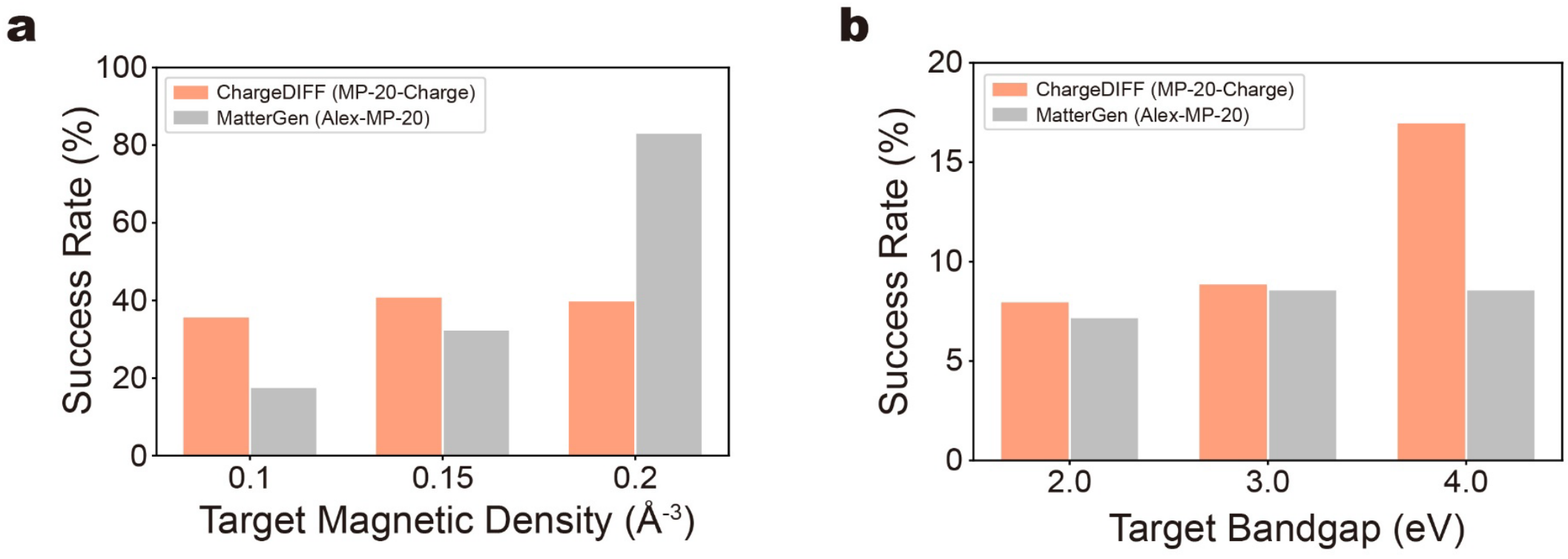

**Fig. S12. Conditional Generation Performance Comparison with MatterGen. a-b.** Conditional generation performance of ChargeDIFF compared with MatterGen for magnetic density (**a**), and bandgap (**b**). For MatterGen, the shared conditional model checkpoint was used without training a new model. The MatterGen model was originally trained in an unconditional manner on the much larger Alex-MP-20 database (607,634 structures) and later fine-tuned on 605,000 labeled data points for magnetic density and 42,000 labeled data points for bandgap. The success rate, which is defined as the proportion of structures with deviation from the target value within 10 %, was evaluated using 500 generated structures.

## Section S4-4. Conditional Generation on Energy Above Hull

In addition to the conditional generation results targeted magnetic density, bandgap, and crystal density shown in Fig. 4 of the main text, conditional generation targeting the energy above hull was additionally performed. Unlike other properties, conditioning on the energy above hull can be jointly applied in various applications to generate more stable materials.

The distributions of energy above hull for 1,000 structures generated with target values of 0.0, 0.05, and 0.1 eV/atom are shown in Fig. S13a. It was observed that structures generated with lower target energy above hull values tended to be distributed toward lower-energy regions. To evaluate the model more quantitatively, the proportion of stable structures (i.e., energy above hull $\leq$ 0.1 eV/atom) among the generated structures was assessed to determine whether the model can effectively control the fraction of stable structures through energy above hull conditioning. As a result, as the target energy above hull decreased from 0.1 to 0.05 to 0.0 eV/atom, the proportion of stable structures increased to 38.2%, 46.8%, and 59.5%, respectively (Fig S13b). This indicates that energy above hull conditioning successfully controls the stability of the generated materials in accordance with the target values.

Similar to Fig. 4g of the main text, an ablation study on charge density was conducted by evaluating a variant model that does not consider charge density (ChargeDIFF (AXL) in Fig. S13b) using the same metrics. While this model also exhibited variations in the stable structure ratio with different target values, its overall ability to generate stable structures across all target ranges was lower than that of the original ChargeDIFF. This is because the charge density plays a crucial role in determining material energies, as in DFT calculations, and, consistent with the results in the main text, demonstrates that incorporating charge density significantly enhances the conditioning capability of ChargeDIFF.

a

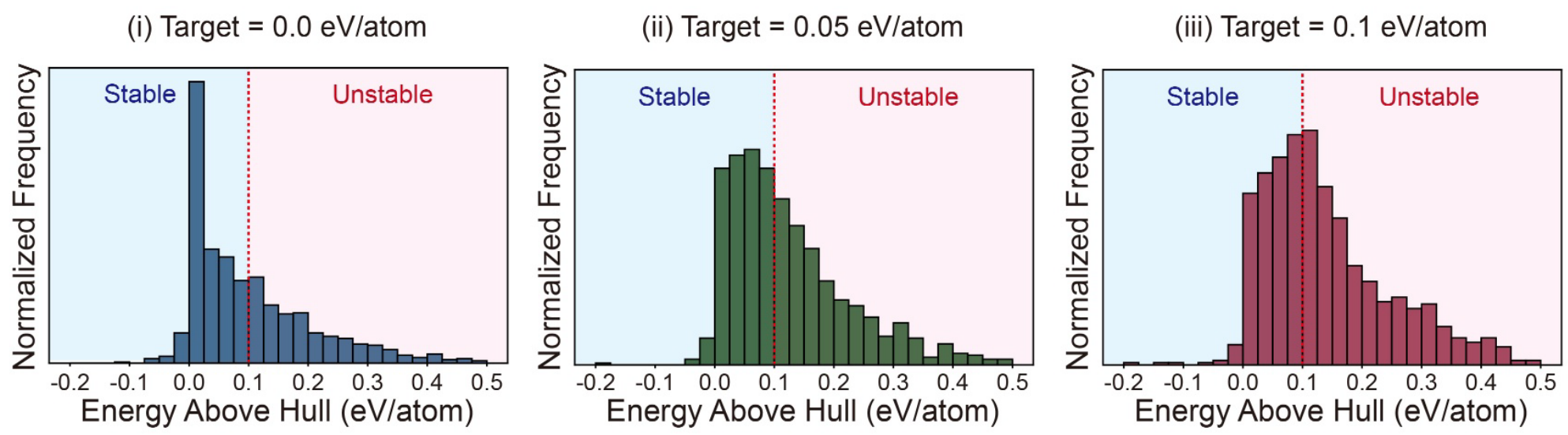

b

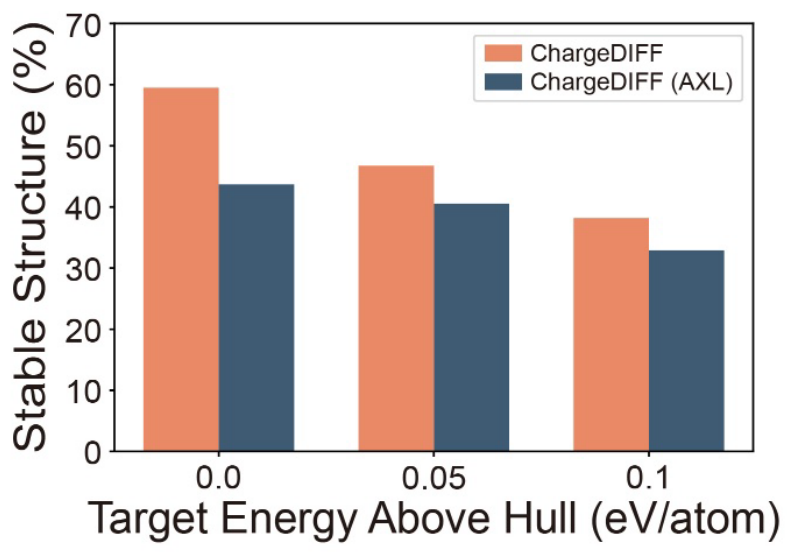

**Fig. S13. Conditional Generation Results for Energy Above Hull. a.** Conditional generation results for energy above hull. The distribution of 1,000 structures generated for each target energy above hull value of 0.0, 0.05, and 0.1 eV/atom is shown. **b.** Ablation study on charge density. Conditional generation results for ChargeDIFF (orange) and its ablated variant without charge density, ChargeDIFF (AXL) (blue), are compared. The comparison was based on the proportion of generated structures that are stable ($E_{hull} \leq 0.1$ eV/atom). This metric reflects how well the conditional model can control the stability of the generated materials according to the desired target energy above hull.

## Section S4-5. Conditional Generation on Space Group

As all crystal structures can be classified into 230 space groups, the space group constitutes a categorical property. The distribution of space group of the structures in the training dataset can be seen in Fig. S2. As an example of conditional generation for categorical data, conditional model for space group was trained and evaluated. Fm/3m, C2/m, and I4/mmm were tested as generation targets, which are top three most occurring space groups among the dataset. 500 structures were generated for each space group and minimal geometry optimization was performed using MatterSim prior to the space group determination. Fig. S14a shows the proportion of the structures that correctly match the desired space group, where 81.8%, 27.0%, and 67.8% of the structures were correctly generated for Fm/3m, C2/m, and I4/mmm, respectively. Considering there are a total of 230 categories, this represents a high accuracy. Furthermore, for C2/m, which exhibited relatively lower accuracy, many structures were classified as Cm, the same monoclinic space group with slightly lower symmetry. Among the generated structures, 60.8% belonged to either of these two space groups. This suggests that the accuracy could be further improved by performing a more rigorous structure optimization (e.g., via DFT) or by appropriately adjusting the tolerance used for space group determination. The sample generated structures as a result of space group conditioning can be found in Fig. S14b-d.

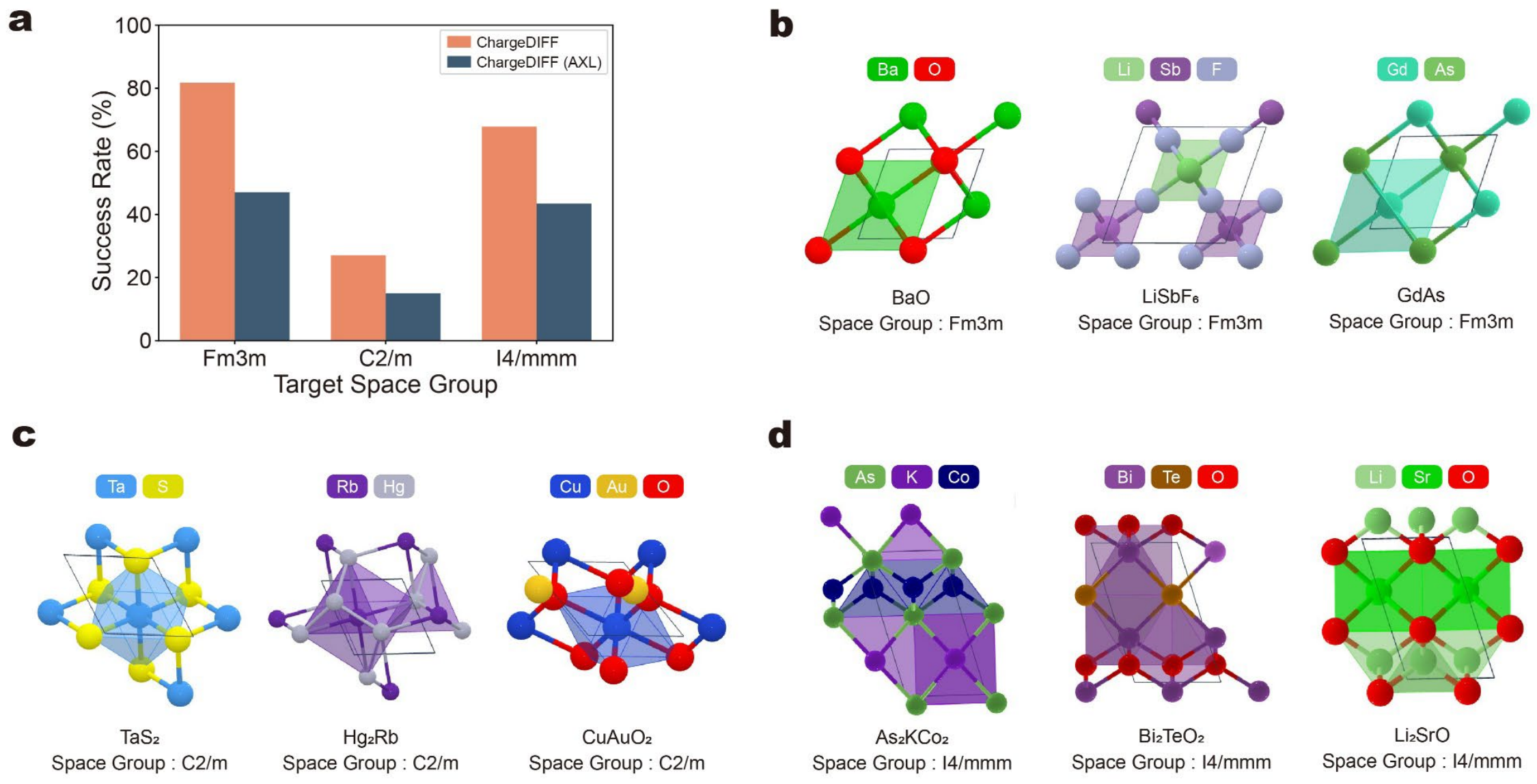

**Fig. S14. Conditional Generation Results for Space Group. a.** Conditional generation results for space groups obtained from ChargeDIFF (orange) and its ablated variant without charge density, ChargeDIFF (AXL) (blue). For the three most common space groups in the database (Fm3m, C2/m, and I4/mmm), 500 structures were generated for each, and the proportion of structures matching the target space group is depicted. **b-d.** Example generated structures from ChargeDIFF with target space groups of Fm3m (**b**), C2/m (**c**), and I4/mmm (**d**).

# Section S5. Details on 3D Charge Density-based Inverse Design

## Section S5-1. Generation and Analysis Scheme

This section provides a detailed explanation of the 3D charge density-based inverse design introduced in the main text and Fig. 5. For the inpainting technique, we adopted the blended diffusion approach proposed by Avrahami et al. Their method involved blending the progressively generated noisy images with the noisy version of the original image in the masked region, thereby producing a seamless result. Following this approach, we added noise to the voxels initialized to zero and blended them with the generated noisy samples within the specified region, thus generating structures in which the charge density of the desired region remains low. To obtain structures with 1D Li ion pathways, the region of x: [0.33, 0.66], y: [0.33, 0.66], z: [0,1] was designated as low charge density region, whereas for 2D pathways, the x: [0, 1], y: [0, 1], z: [0.33, 0.66] was designated.

Conditioning on the chemical system is performed via multi-hot encoding as shown in Fig. S10. Unlike the conventional approach, where each atom type is treated as a separate class in multi-hot encoding to generate materials with specific atom combinations, we constrain the generation to a class of materials, such as transition metal oxides, while allowing flexibility in the choice of metal species. To achieve this, we leveraged group-level multi-hot encoding (rather than atom-level), and the grouping of atoms used in this work can be found in Fig. S16. Information about which group each material belongs to is incorporated during training, and in the generation phase, the model is capable of generating materials from the target group (i.e., target material class). In this work, the generation of cathode materials was based on combinations of $4^{th}$-period transition metals and oxygen, enabling the generation of transition metal oxides encompassing a variety of metal species.

After generating materials via inpainting and chemical system conditioning, the structures were optimized using DFT, followed by the insertion of Li ions using the code from Shen et al. (Fig. S15). In previous work, they reported that Li ions tend to be inserted into regions of low charge density within cathode materials, and shared a python code for this task. Using this code, we inserted Li ions to generated transition metal oxides, which are presented as LIB cathode materials with desired ion migration pathways.

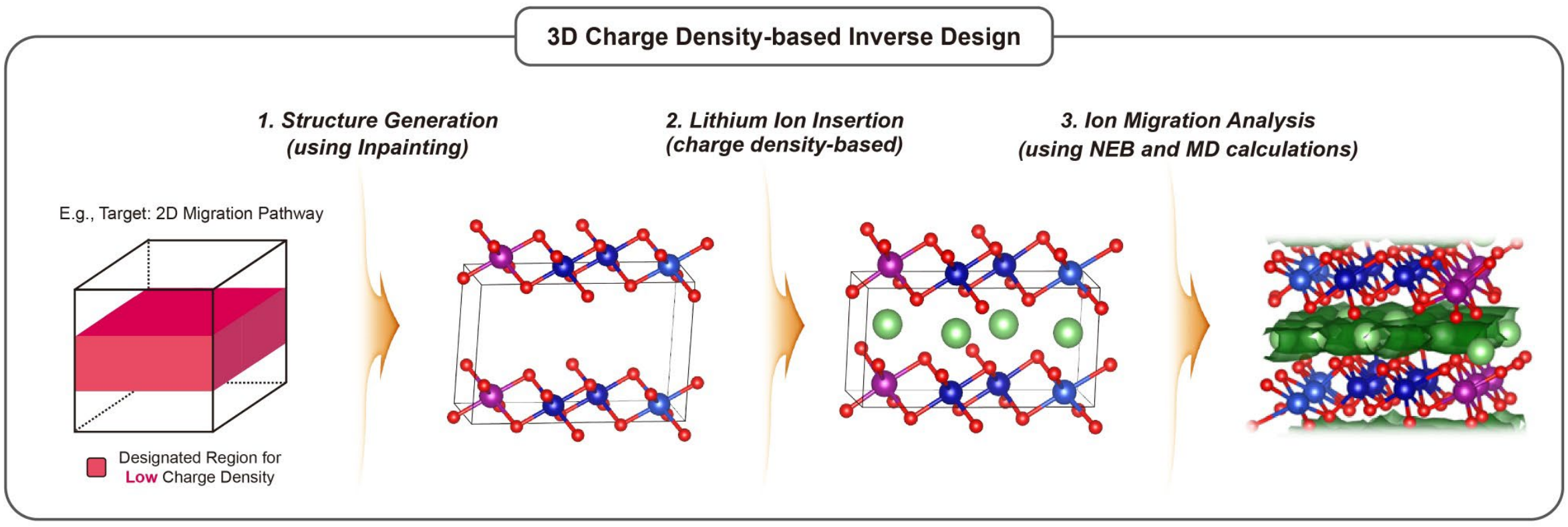

**Fig. S15. 3D Charge Density-based Inverse Design Scheme.** Schematic describing the generation process of LIB cathode materials using 3D charge density-based inverse design. First, structures are generated by specifying the desired low charge density regions using an inpainting technique. Next, lithium ions are inserted into the structures using the code from Shen et al. The resulting structures are then analyzed for ion migration pathways using both topological analysis and MD simulations.

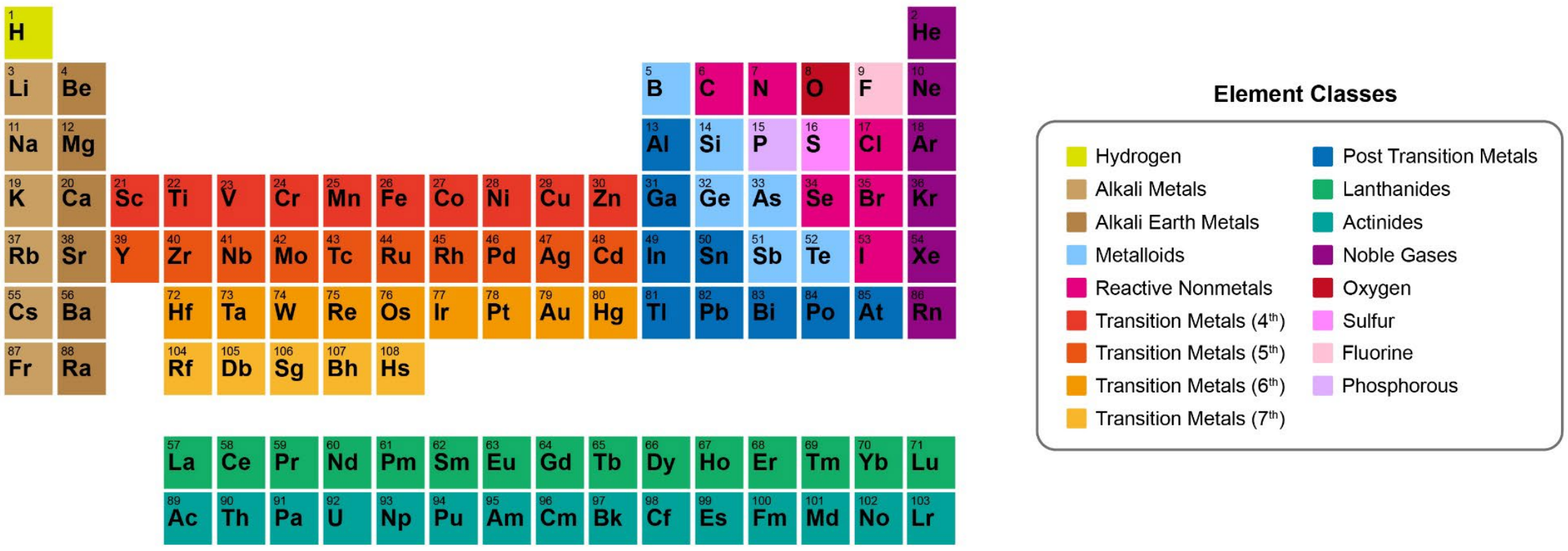

**Fig. S16. Grouping of Atoms for Chemical System Conditioning.** To enable group-level multi-hot encoding of chemical system conditioning, the atoms in the periodic table were classified into 17 groups. For the generation of transition metal oxide cathode materials, multi-hot embeddings of transition metal ($4^{th}$ period) and oxygen were used in this work. This grouping can be made more detailed or coarser depending on the application.

## Section S5-2. Inverse Design Results

Using the charge density-based inverse design approach described above, we generated 1,000 structures each with 1D and 2D target migration pathways, and subsequently analyzed them (Fig. S17a). Initially aiming for transition metal oxides, we discarded structures containing non-target elements or consisting solely of oxygen of metals. Afterwards, we checked whether one or more Li ions were successfully inserted into the structures. For the remaining structures, the dimensionality of Li-ion channels was determined through a network analysis based on Li-Li distances with a 4 Å cutoff. As a result, a total of 323 and 413 structures were found to possess 1D and 2D migration channel dimensionality, respectively. Given that no prior research has explored inverse design for ion migration pathways, these results indicate a high success rate and confirm that our proposed inverse design approach can produce cathode materials with the targeted channel topologies.

Beyond analyzing channel dimensionality via network analysis, additional nudged elastic band (NEB) calculations were conducted to assess the potential of the generated structures as cathode materials. Given the large number of candidates, a further down-selection was performed. (Note that this procedure was not intended to filter structures based on success or failure, but rather to reduce the number of structures for NEB calculations.) We first analyzed the oxygen coordination of each metal and Li ion, selecting structures in which they exhibit only 4- or 6- coordination, the most common coordination. In addition, we selected only the SUN structures, resulting in a final set of 32 structures with 1D pathways and 14 structures with 2D pathways. For these structures, the energy barriers were evaluated using NEB calculations (detailed in Methods section of main text).

NEB calculations were successfully performed for 28 and 10 structures (respectively for 1D and 2D), and the distribution of their energy barriers is shown in Fig. S17b. The energy barriers were evaluated using a 0.5 eV threshold, which was arbitrarily chosen cutoff based on simulation results for existing commercial cathode materials (LCO: 0.647 eV, LFP: 0.235 eV). As a result, 9 structures with 1D pathways and 10 structures with 2D pathways met this criterion (Fig S18, S19), exhibiting energy barriers comparable to those of practical cathode materials. These findings suggest that the inverse design approach proposed in this study can successfully generate cathode materials with practical feasibility.

While inpainting-based materials design demonstrate promising potential, as demonstrated in the present study, it is expected to undergo further refinement moving forward. Currently, it primarily employs a conventional masked inpainting strategy, in which a hard replacement of the masked region is performed. In some cases, this can lead to unrealistic structural distortions in the regions surrounding the masked channels. As an alternative, approaches such as soft inpainting and CFG, which allows for smoother integration of the inpainted regions with their surroundings, could be considered to mitigate these artifacts. Inpainting-based generation represents an area with considerable room for advancement, and a variety of techniques could potentially be applied to generate more seamless structures while satisfying the desired guidance.

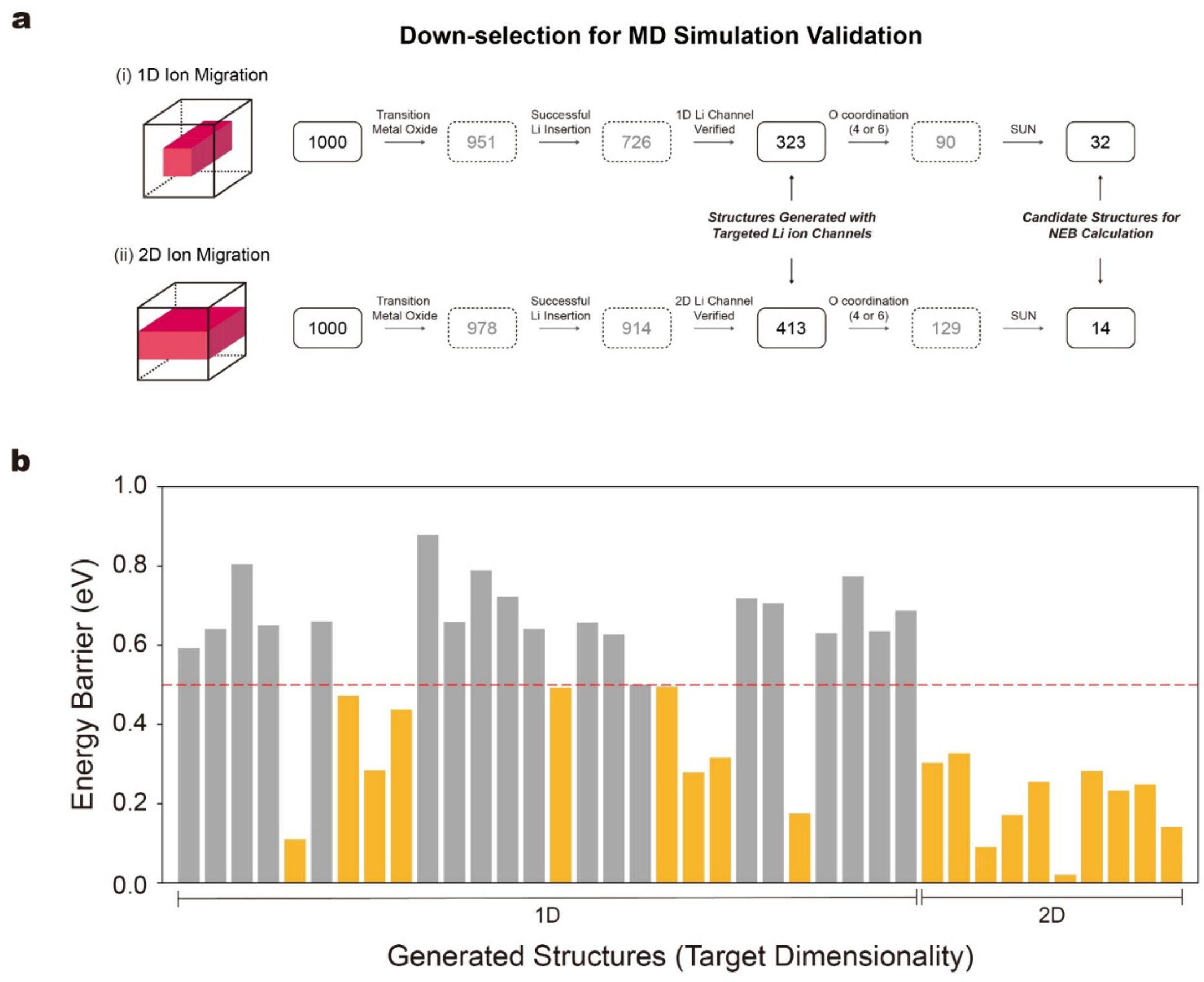

**Fig. S17. 3D Charge Density-based Inverse Design Results. a.** Evaluation and analysis of cathode materials generated via charge density-based inverse design. A total of 1,000 structures were generated targeting 1D and 2D ion migration channels. Each structure was evaluated to determine whether the target dimensionality channel is present based on: (1) whether it is a transition metal oxide, (2) whether at least one Li ion could be successfully inserted using the insertion code, and (3) whether the topological analysis confirms the presence of the desired Li ion channels. Subsequently, structures were further down-selected for NEB calculations to determine the energy barriers for ion migration. The selection was based on (1) whether each metal and Li ion forms 4- or 6-fold coordination with oxygen, and (2) whether the structure is SUN (stable, unique and new). **b.** Distribution of energy barrier values obtained from NEB calculations. The energy barriers for structures selected in (a) that successfully completed MD simulations are shown, using 0.5 eV as a reference threshold. This threshold was set with reference to energy barrier values of practically used cathode materials under the same settings. (LCO: 0.647 eV, LFP: 0.235 eV)

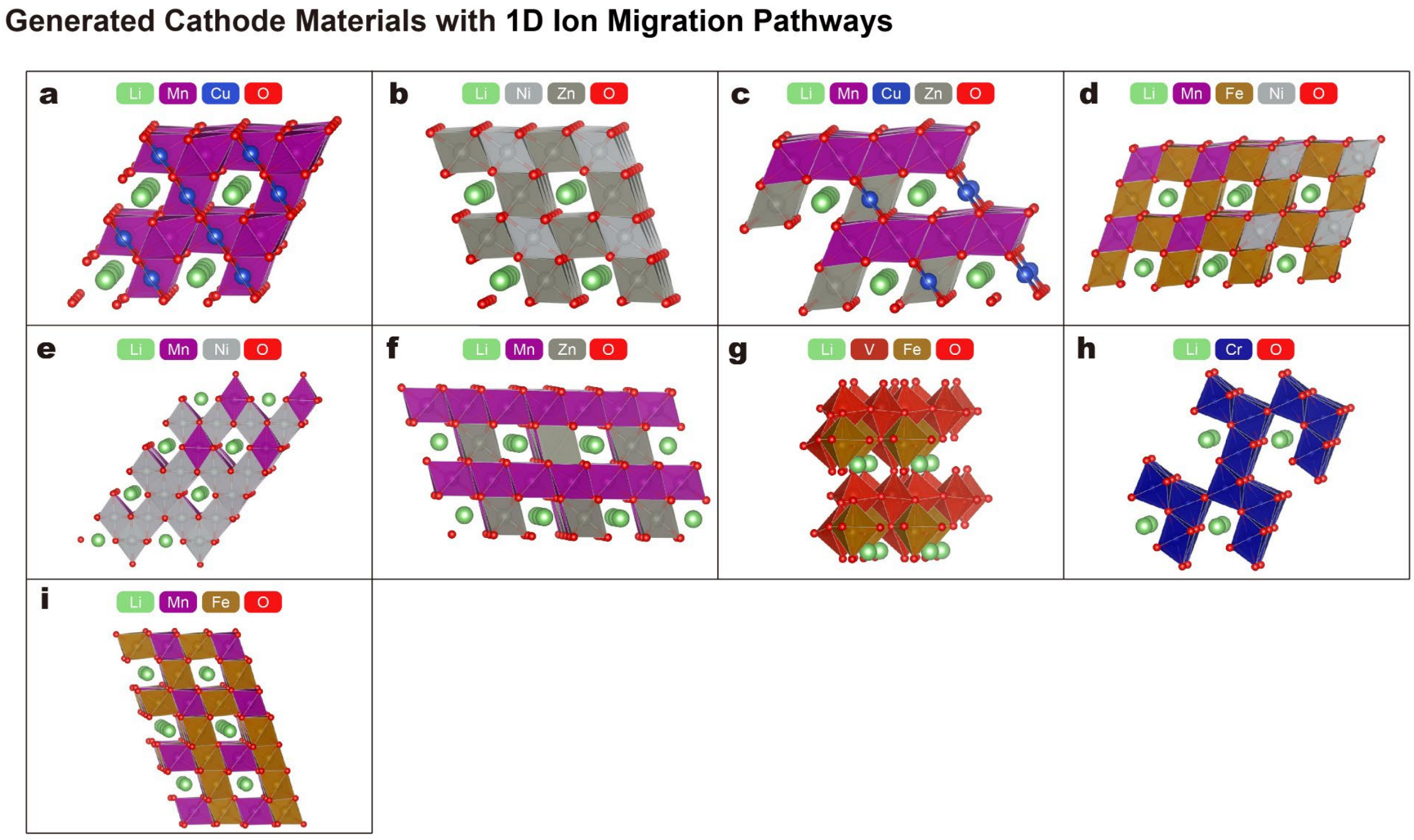

**Fig. S18. Cathode Materials with 1D Ion Migration Pathways Generated via Charge Density-based Inverse Design.** From the set of structures generated through charge density-based inverse design targeting a 1D ion migration pathway, atomic structures of those exhibiting a Li-ion migration energy barrier of less than 0.5 eV (highlighted in Fig. S17b) are shown. Structures are displayed as 2×2×2 supercell, with the cell lattice omitted for clarity.

**Generated Cathode Materials with 2D Ion Migration Pathways**

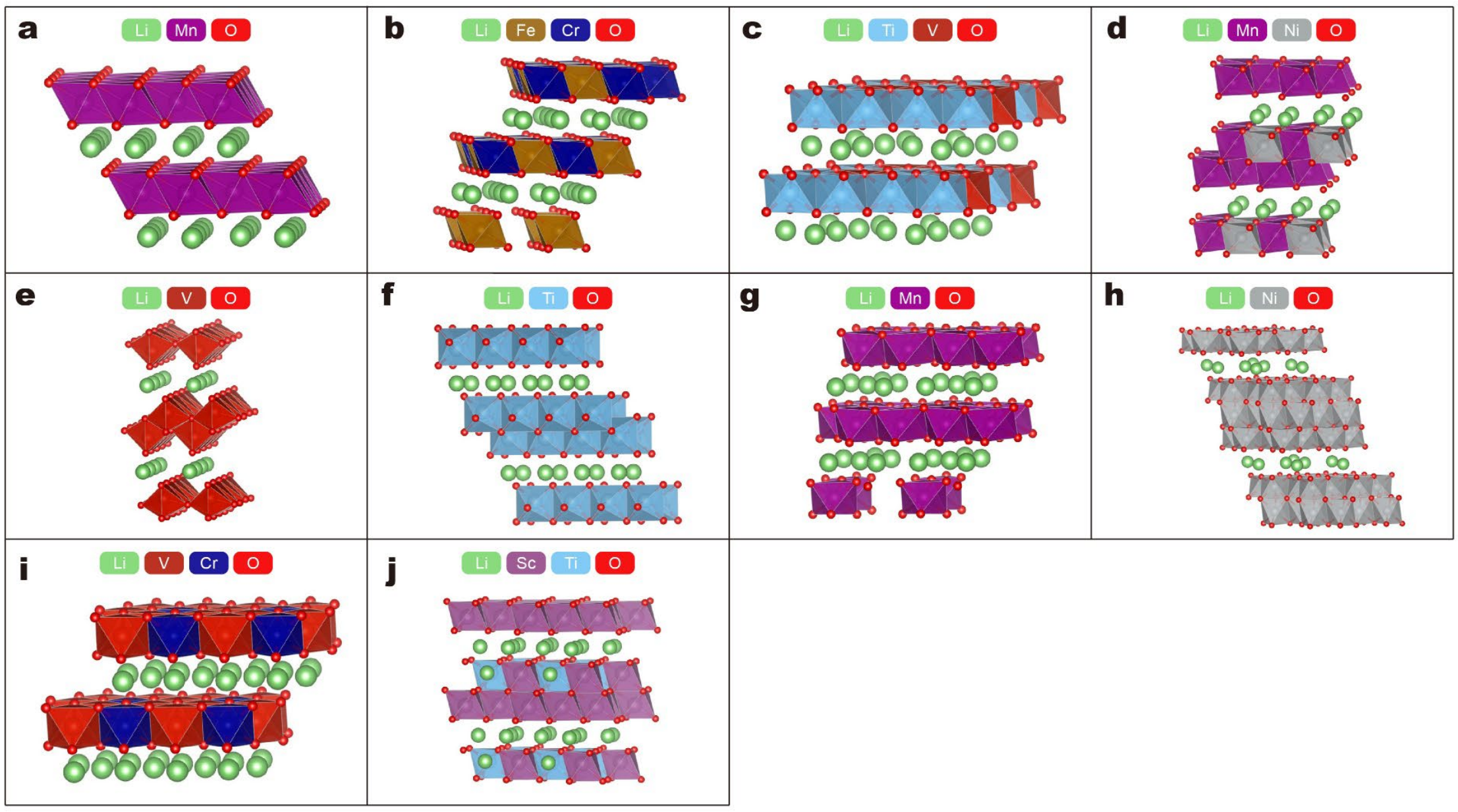

**Fig. S19. Cathode Materials with 2D Ion Migration Pathways Generated via Charge Density-based Inverse Design.** From the set of structures generated through charge density-based inverse design targeting a 2D ion migration pathway, atomic structures of those exhibiting a Li-ion migration energy barrier of less than 0.5 eV (highlighted in Fig. S17b) are shown. Structures are displayed as 2×2×2 supercell, with the cell lattice omitted for clarity.